\documentclass[prd,
preprint,nofootinbib%
 ,secnumarabic%
,amssymb, amsmath,mathrsfs,nobibnotes,aps,11pt]{revtex4}
\pdfoutput=1
\usepackage{bm}%
\usepackage{epsfig}
\usepackage{color}
\usepackage{ulem}
\usepackage{graphicx}
\usepackage{float}
\usepackage[caption=false]{subfig}
\usepackage{braket}
\usepackage{slashed}
\usepackage{subfig}
\usepackage{comment}

\usepackage{tabularx}
\usepackage[colorlinks=true,citecolor=blue,linkcolor=blue]{hyperref}%
\usepackage{ragged2e}
\newcommand{\PlotPairWide}[4]{%
\begin{figure}[htp]
\centering
\hspace{-5pt}
\includegraphics[width=3.1in]{#1}
\hspace{+5pt}
\includegraphics[width=3.1in]{#2}
\begin{minipage}[l]{0.8\linewidth}\vspace{-10pt} \justifying \small {\raggedleft \caption{#4} \label{#3}}\end{minipage}\vspace{0pt}
\end{figure}}

\begin{document}
%

\begin{flushright}
\begin{tabular}{c c c c c}
OSU-HEP-17-03 &  & MI-TH-17637 &  & UH-511-1279-2017
\end{tabular}
\end{flushright}

\title{Probing Squeezed Bino-Slepton Spectra with the Large Hadron Collider}

\author{Bhaskar Dutta$^{1,}$\footnote{ dutta@physics.tamu.edu }}
\author{Kebur Fantahun $^{2,}$\footnote{  kxf010@shsu.edu }}
\author{Ashen Fernando$^{2,}$\footnote{ baf024@shsu.edu }}
\author{Tathagata Ghosh$^{3,}$\footnote{tghosh@okstate.edu}}
\author{Jason Kumar$^{4,}$\footnote{jkumar@hawaii.edu}}
\author{Pearl Sandick$^{5,}$\footnote{ sandick@physics.utah.edu }}
\author{Patrick Stengel$^{6,}$\footnote{ pstengel@umich.edu }}
\author{Joel W. Walker$^{2,}$\footnote{ jwalker@shsu.edu }}
\affiliation{$^1$Department of Physics \& Astronomy, Texas A\&M University, College Station, TX 77843, USA}
\affiliation{$^2$Department of Physics, Sam Houston State University, Huntsville, TX 77341, USA}
\affiliation{$^3$Department of Physics, Oklahoma State University, Stillwater, OK 74078, USA}
\affiliation{$^4$Department of Physics \& Astronomy, University of Hawaii, Honolulu, HI 96822, USA}
\affiliation{$^5$Department of Physics and Astronomy, University of Utah, Salt Lake City, UT 84112}
\affiliation{$^6$Michigan Center for Theoretical Physics, Department of Physics, University of Michigan, Ann Arbor, MI 48109, USA.}

\def\be{\begin{equation}}
\def\ee{\end{equation}}
\def\al{\alpha}
\def\bea{\begin{eqnarray}}
\def\eea{\end{eqnarray}}

\def\tev{\, {\rm TeV}}
\def\gev{\, {\rm GeV}}
\def\mev{\, {\rm MeV}}
\def\kev{\, {\rm keV}}
\def\swsq{\sin^2\theta_W}
\def\xfb{\,~{\rm fb}}
\newcommand{\sigmaSI}{\sigma_{\rm SI}}
\newcommand{\sigmaSD}{\sigma_{\rm SD}}
\newcommand{\gsim}{\lower.7ex\hbox{$\;\stackrel{\textstyle>}{\sim}\;$}}
\newcommand{\lsim}{\lower.7ex\hbox{$\;\stackrel{\textstyle<}{\sim}\;$}}
\newcommand{\fb}{\rm fb}
\newcommand{\ifb}{\rm fb^{-1}}
\newcommand{\pb}{\rm pb}
\newcommand{\ipb}{\rm pb^{-1}}
\newcommand{\m}{\rm m}
\newcommand{\ab}{\rm ab}

\newcommand{\drawsquare}[2]{\hbox{%
\rule{#2pt}{#1pt}\hskip-#2pt
\rule{#1pt}{#2pt}\hskip-#1pt
\rule[#1pt]{#1pt}{#2pt}}\rule[#1pt]{#2pt}{#2pt}\hskip-#2pt
\rule{#2pt}{#1pt}}

\newcommand{\fund}{\raisebox{-.5pt}{\drawsquare{6.5}{0.4}}}
\newcommand{\Ysymm}{\raisebox{-.5pt}{\drawsquare{6.5}{0.4}}\hskip-0.4pt%
        \raisebox{-.5pt}{\drawsquare{6.5}{0.4}}}
\newcommand{\Yasymm}{\raisebox{-3.5pt}{\drawsquare{6.5}{0.4}}\hskip-6.9pt%
        \raisebox{3pt}{\drawsquare{6.5}{0.4}}}
\newcommand{\antifund}{\overline{\fund}}
\newcommand{\bYasymm}{\overline{\Yasymm}}
\newcommand{\bYsymm}{\overline{\Ysymm}}
\newcommand{\Dsl}[1]{\slash\hskip -0.20 cm #1}

\newcommand{\ssection}[1]{{\em #1.\ }}

\declareslashed{}{/}{0}{-0.05}{E}
\def\met{{\slashed{E}} {}_{T}}

\begin{abstract}
We consider a Minimal Supersymmetric Standard Model
scenario in which the only light superparticles are a bino-like dark matter
candidate and a nearly-degenerate slepton.  It is notoriously difficult to probe this scenario
at the Large Hadron Collider, because the slepton pair-production process yields a final state with soft leptons and
small missing transverse energy.  We study this scenario in the region of parameter space where the mass difference
between the lightest neutralino and the lightest slepton ($\Delta m$) is $\lesssim 60~{\rm GeV}$,
focusing on the process in which an additional radiated jet provides a transverse boost to the slepton
pair. We then utilize the angular separation of the leptons from
each other and from the missing transverse energy, as well as the angular separation between the jet
and the missing transverse energy, to distinguish signal from background events.
We also use the reconstructed ditau mass, the $\cos \theta^*_{\ell_1 \ell_2}$ variable,
and for larger $\Delta m$, a lower bound on the lepton $p_T$.
These cuts can dramatically improve both signal sensitivity and the signal-to-background ratio, permitting
discovery at the Large Hadron Collider with reasonable integrated luminosity over the interesting region of
parameter space. Using our search strategy the LHC will be able to exclude
$m_{\tilde{\mu}} \approx 200$ GeV for $\Delta m \lesssim 60$ GeV at $1.5-3 \sigma$ with 1000 fb$^{-1}$ of integrated luminosity.
Although we focus on a particular model, the results generalize to a variety of scenarios
in which the dark matter and a leptonic partner are nearly degenerate in mass, and especially 
to scenarios featuring a scalar mediator.

\end{abstract}
\maketitle

\section{Introduction}

A well-studied scenario for physics beyond the Standard Model (SM) is one in which there is a new spin-0 particle ($\tilde \ell$)
with the same Standard Model gauge quantum numbers as a lepton, and a spin-1/2 dark matter particle  ($\chi$)
which is a Standard Model singlet.  This scenario arises in a variety of specific models of new
physics, including the MSSM, in the case where the lightest neutralino is bino-like; in this case, $\tilde \ell$ is
a slepton and $\chi$ is the bino.  This scenario
has thus been the subject of intense study at the Large Hadron Collider (LHC).

The standard LHC strategy for probing this scenario is to search for ${\tilde \ell}^* \tilde \ell$
pair-production, with each $\tilde \ell$ decaying promptly to $\chi$ and a Standard Model lepton ($\ell$).
This process yields  a distinctive signature: two opposite-sign leptons and missing transverse energy.  But this
strategy tends to fail in the region of parameter space where $\tilde \ell$ and $\chi$ are nearly mass-degenerate
($\Delta m \equiv m_{\tilde \ell}- m_\chi \lesssim  60\gev$)~\cite{Aad:2014vma,Khachatryan:2014qwa},
because (i) the $\tilde \ell$ are typically produced with
relatively small momenta, implying that the lepton momenta and the missing transverse energy are small in the nearly-degenerate
limit, and (ii) processes with equivalent final state topology, i.e.~$\bar t t$ and $V,VV+{\rm jets}$ (where $V$ is an electroweak
gauge boson), become a major source of background since the leptons arising from $W,Z$-decays have $p_T$ of around $40 \gev$.

The ATLAS collaboration has searched~\cite{ATLAS:2017uun} for electroweak production of sleptons with decay to
neutralino plus lepton at the LHC13 ($\mathcal{L}=36.1~{\rm fb}^{-1}$) in the zero-jet dilepton final state.
Slepton masses up to 500 GeV are excluded, if one assumes a massless neutralino.  However, this search is
generally insensitive to the narrow mass splitting $\Delta m \lesssim 60$~GeV regime.
CMS has searched~\cite{CMS:2017fij} for soft opposite-sign leptons at 13 TeV with $\mathcal{L}=35.9~{\rm fb}^{-1}$,
focusing on the pair-production of charginos and neutralinos $\chi^\pm_1,\chi^0_2$ with nearly degenerate
mass.  This study excludes the parameter space up to around 230~GeV, for mass gaps from the lightest
neutralino as small as 20~GeV, but no slepton limits are inferred.

Attempts were made to understand this nearly degenerate region for $\Delta m\leq 20$ GeV,
utilizing the Vector Boson Fusion topology~\cite{Dutta:2014jda} and monojet plus dileptons~\cite{Han:2014aea,Barr:2015eva},
and it was found that an upper bound on the lepton transverse momentum ($p_T<30\gev$) is useful for investigating
this region at the LHC.
Ref.~\cite{Han:2014aea} also made substantial use of the kinematic variable $M_{T2}$,
which sets an upper bound on the mass of pair-produced parent particles that decay into
the two visible systems (after making a specific hypothesis for the mass of the associated invisible species).
The presence of the  jets in both search strategies gives a transverse
boost to the $\tilde \ell^* \tilde \ell$ system, which increases the lepton momenta and the
missing transverse energy in order to reduce the SM background. However, none of these search strategies
work for $\Delta m\sim 25-60 \gev$, due to  the aforementioned background $W,\,Z$ decay processes.

In this work, we consider supplementary strategies for probing this nearly degenerate region of parameter space utilizing
searches for a single jet, in addition to the opposite-sign dilepton and missing
transverse momentum.
We show that the $VV$ background can be reduced significantly, and the signal to background ratio elevated,
through additional cuts based on the angular
distribution of the leptons and $\met$ and the azimuthal angular separation between the jet and the $\met$.
These new selection processes allow us to provide an alternative complementary formulation for the
investigation of the very low mass splitting regime ($\Delta m\sim 10$~GeV), and also to cope
with slightly larger splittings, up to about $60$~GeV.
We will show that models with $\Delta m \leq 60\gev$ can be probed at $4-9\sigma$ confidence
with reasonable luminosity (around $300~{\rm fb}^{-1}$) at the LHC for smuon masses around 110~GeV, and will further estimate the efficacy of our analysis
for benchmarks with heavier smuons.

The nearly-degenerate region of parameter space has been of great interest due to the novel  early
Universe cosmology which is possible for such models.  For example, if there are non-minimal flavor-violating
couplings and if $m_{\chi} \sim  m_{\tilde \ell} \sim {\cal O}(100)\gev$, then dark matter annihilation
($\chi \chi \rightarrow \bar \ell \ell$) through the $t$-channel
exchange of $\tilde \ell$ can deplete the dark matter relic density enough to achieve
consistency with observation~\cite{bulk, Fukushima:2014yia}.  If $\Delta m$ is sufficiently small ($\leq 60$\gev), then both
$\chi$ and $\tilde \ell$ may be abundant at the time of dark matter freeze-out, and co-annihilation may also
be important for depleting the dark matter relic density~\cite{coannihilation, Coann}.  It is thus quite interesting to develop collider
tools for probing this region of parameter space.

The structure of this paper is as follows.  In section 2, we describe the underlying strategy for
searching for models with a nearly degenerate bino-slepton pair.  In section 3, we describe the optimization 
of selection cuts for different mass-splitting scenarios, and determine the signal significance and signal-to-background ratio the LHC could provide,
for a variety of benchmark choices.  We conclude in section 4 with a discussion of our results.

\section{Strategy}

We consider a simplified scenario in which the only light sparticles are a single slepton
($\tilde \ell$) and a bino ($\chi$), and the only allowed decay for the slepton is
$\tilde \ell \rightarrow \chi \ell$.  Although we have described this scenario in the language
of the MSSM, the same scenario arises in a variety of models for new physics (including, for
example, WIMPless dark matter~\cite{Feng:2008ya,Feng:2008mu,Barger:2010ng,Fukushima:2011df}).

The standard strategy for probing this scenario at the LHC is to search for the process
$\bar q q \rightarrow \gamma^*/Z^* \rightarrow \tilde \ell^* \tilde \ell \rightarrow \chi \chi \bar \ell \ell$.
The signature for this process is the production of a same-flavor, opposite-sign dilepton pair, accompanied
by missing transverse energy ($\met$). But for most signal events, the $\tilde \ell^* \tilde \ell$ pair are
produced with negligible transverse boost.  If $\Delta m = m_{\tilde \ell} - m_\chi$ is small, then the $\chi$ produced by
$\tilde \ell/\tilde \ell^*$ decay is non-relativistic and the missing transverse energy is small; in this
case, the signal is indistinguishable from the large background arising from the Drell-Yan process
$\bar q q \rightarrow \gamma^*/Z^* \rightarrow \bar \ell \ell$.  As a result, this LHC search strategy
is largely insensitive for $\Delta m \lesssim 60\gev$.

To probe this region of parameter space, we will instead focus on the process
$g q \, (\bar q q) \rightarrow j\gamma^*/Z^* \rightarrow j\tilde \ell^* \tilde \ell \rightarrow j\chi \chi \bar \ell \ell$,
wherein a single additional hard jet is emitted by one of the initial gluon or quarks of the hard process.  The emission
of a hard jet gives a large transverse boost to the $\tilde \ell^* \tilde \ell$ system; the decay products of this
system are now collimated, and the $\chi \chi$ pair can have a significant transverse momentum which appears as
$\met$~\cite{ATLAS:2016,CMS:2017fij}.

\subsection{Leading Backgrounds and Primary Event Selections}

The dominant SM background processes are:
\begin{itemize}
\item{$pp \rightarrow jZ \rightarrow j \bar \tau \tau \rightarrow j \bar \ell \ell \bar \nu \nu \bar \nu_\tau \nu_\tau$,}
\item{$pp \rightarrow \bar t t (j) \rightarrow \bar b W^- b W^+  (j)\rightarrow \bar b b \bar \ell \ell \bar \nu \nu (j)$,}
\item{$pp \rightarrow jZZ/jW^+W^- \rightarrow j\bar \ell \ell \bar \nu \nu, j \bar \ell \ell \bar \tau \tau (\rightarrow
\bar \ell \ell \bar \nu \nu + {\rm jets})$}
\end{itemize}
where in all cases, the missing transverse energy arises from the neutrinos.

Since the background processes will contribute equally to dimuon and dielectron final states,
whereas the signal may be distinguished based upon the identity of the slepton, we opt here to
consider $\ell = \mu$ and $\tilde \ell = \tilde \mu$ (smuon)\footnote{In particular, we will assume 
$\tilde \ell = \tilde \mu_L$, although this choice will have little effect on cut selection.}.  The electron and selectron
scenario will be essentially identical, with some differences emerging at the detector level.
The muon has certain advantages in identification, associated with secondary observation in the dedicated exterior detector systems.
Since the decay of $WWjj$ to opposite-sign dileptons is expected to produce $(e^+e^-:\mu^+\mu^-:e^\pm\mu^\mp)$
in the ratio $(1:1:2)$, an interesting parallel strategy involves leveraging the differential measurement of pure
and/or mixed flavor dileptons, as well as associated differential kinematic distributions,
in order to estimate and control background~\cite{Baer:2014kya}; however, we will not pursue this possibility further in the current work.
The minimal muon transverse momentum is identified in
our default efficiency formula as 10~GeV, although we will subsequently investigate the relaxation of that parameter.
For specificity, we consider $m_{\tilde \mu} = 110\gev$ for the main analysis, but likewise subsequently
investigate the reach of our analysis with heavier smuon masses.
Correspondingly, our main analysis will focus on six benchmark scenarios, variously
with neutralino masses of  $m_\chi = 50, 60, 70, 80, 90,~{\rm and}~100$~GeV.

We will define ``primary'' cuts as those which are imposed prior to the analysis,
for the purpose of fundamentally characterizing the targeted dimuon plus boosted ISR jet event topology.
We take a lead at this level from Refs.~\cite{Han:2014aea, Barr:2015eva} and~\cite{Baer:2014kya},
which draw in turn from Ref.~\cite{Han:2014kaa}.
In addition to requiring an opposite-sign dimuon pair, we require one and only one jet with $P_T^{j1} > 30$~GeV,
and enforce a nominal lower bound on missing transverse energy of $\met > 30$~GeV
(hardness of the single jet and the missing momentum will both be substantially escalated at the stage of secondary optimization).
We also veto on $b$-tagged jets and tagged hadronically decaying taus, which are reconstructed at detector level.
These last cuts significantly reduce the background from
$pp \rightarrow \bar t t$  and from SM processes in which $\tau$s decay hadronically.
These cuts are summarized in Table~\ref{tab:PrimaryCuts}, along with the residual effective LHC14 cross sections
for the $\bar{t}t+$Jets, $\tau\tau+$Jets,  $Z+$Jets, and $VV+$Jets background components,
as well as each of the six signal benchmark models.
In what follows, all event shape distributions will be shown after the imposition of these primary cuts.

\bgroup
\def\arraystretch{1.3}
\begin{table}[!htp]
\caption{Jet matched production and residual effective cross sections (fb) at the LHC14
are tabulated for the $\bar{t}t+$Jets, $\tau\tau$+Jets, $Z+$Jets, and $VV+$Jets backgrounds, as well
as the six signal benchmarks $S_\Delta^{110}$, with $m_{ \tilde{\mu}} = 110$~GeV, and $m_\chi= (110 - \Delta)$~GeV.
These primary cuts are related to the targeted dilepton plus boosted ISR jet event topology and are applied to all events.\\}
\label{tab:PrimaryCuts}
\scalebox{1.11}{%
\begin{tiny}
\hspace{-40pt}
\begin{tabular}{|c||c|c|c|c|c|c||c|c||c|c||c|c|}
\hline
Selection & $t\bar{t}jj$ & $\tau\tau jj$ & $Zjjjj$ & $ZZjj$ & $WZjj$ & $WWjj$ & $S^{110}_{10}$ & $S^{110}_{20}$ & $S^{110}_{30}$ & $S^{110}_{40}$ & $S^{110}_{50}$ & $S^{110}_{60}$ \\
\hline
\hline
Matched Production & $6.1 \times 10^{5}$ & $5.6 \times 10^{4}$ & $5.2 \times 10^{7}$ & $1.3 \times 10^{4}$ & $4.2 \times 10^{4}$ & $9.5 \times 10^{4}$ & $1.9 \times 10^{2}$ & $1.9 \times 10^{2}$ & $1.9 \times 10^{2}$ & $1.9 \times 10^{2}$ & $1.9 \times 10^{2}$ & $1.9 \times 10^{2}$ \\
\hline
$\tau$-veto & $5.4 \times 10^{5}$ & $3.0 \times 10^{4}$ & $5.1 \times 10^{7}$ & $1.2 \times 10^{4}$ & $4.0 \times 10^{4}$ & $8.9 \times 10^{4}$ & $1.9 \times 10^{2}$ & $1.9 \times 10^{2}$ & $1.9 \times 10^{2}$ & $1.9 \times 10^{2}$ & $1.9 \times 10^{2}$ & $1.9 \times 10^{2}$ \\
\hline
OSSF muon & $3.5 \times 10^{3}$ & $4.3 \times 10^{2}$ & $6.0 \times 10^{5}$ & $3.2 \times 10^{2}$ & $5.8 \times 10^{2}$ & $5.1 \times 10^{2}$ & $3.9 \times 10^{1}$ & $6.8 \times 10^{1}$ & $8.1 \times 10^{1}$ & $8.8 \times 10^{1}$ & $8.9 \times 10^{1}$ & $9.1 \times 10^{1}$ \\
\hline
exactly 1J $P_T>30$ & $6.6 \times 10^{2}$ & $2.6 \times 10^{2}$ & $7.1 \times 10^{4}$ & $9.4 \times 10^{1}$ & $1.5 \times 10^{2}$ & $1.1 \times 10^{2}$ & $7.6 \times 10^{0}$ & $1.3 \times 10^{1}$ & $1.6 \times 10^{1}$ & $1.7 \times 10^{1}$ & $1.7 \times 10^{1}$ & $1.8 \times 10^{1}$ \\
\hline
Jet $b$-veto & $1.9 \times 10^{2}$ & $2.5 \times 10^{2}$ & $7.0 \times 10^{4}$ & $8.0 \times 10^{1}$ & $1.4 \times 10^{2}$ & $1.1 \times 10^{2}$ & $7.5 \times 10^{0}$ & $1.3 \times 10^{1}$ & $1.6 \times 10^{1}$ & $1.7 \times 10^{1}$ & $1.7 \times 10^{1}$ & $1.8 \times 10^{1}$ \\
\hline
$\met > 30$~GeV & $1.6 \times 10^{2}$ & $1.8 \times 10^{2}$ & $8.9 \times 10^{3}$ & $3.3 \times 10^{1}$ & $6.6 \times 10^{1}$ & $9.2 \times 10^{1}$ & $6.3 \times 10^{0}$ & $1.0 \times 10^{1}$ & $1.3 \times 10^{1}$ & $1.4 \times 10^{1}$ & $1.5 \times 10^{1}$ & $1.6 \times 10^{1}$ \\
\hline
\end{tabular}
\end{tiny}}
\end{table}
\egroup

For simulation of the signal and backgrounds events, we use {\sc MadGraph5 v2.3.3}~\cite{Alwall:2014hca}
with the {\sc NNPDF23 LO}~\cite{Ball:2012cx} parton distribution function. We pass
our simulated events to {\sc Pythia v6.4}~\cite{Sjostrand:2006za} for showering and hadronization, and subsequently
to {\sc Delphes v3.3 3}~\cite{deFavereau:2013fsa} for detector simulation. All the signal samples, plus ditop, ditau, and
diboson processes are simulated inclusively with up to two partons, while single vector backgrounds are
generated including up to four partons.  The MLM scheme~\cite{MLM} for jet-parton matching has been employed to avoid double counting.
All computations are performed at tree-level, with no $K$-factors included.
Using {\sc PROSPINO 2}~\cite{Beenakker:1996ed}, we estimate an NLO K-factor
of about 1.3 for the 110~GeV signal model, which compares to around
1.3~\cite{Jung:2013vpa} and 1.7 for the sub-leading and leading (after all cuts)
$t\bar{t}$ and $W^+W^-$ backgrounds, respectively.  This does not substantially
affect projected significances ($\sqrt 1.7 \simeq 1.3$).
The {\sc Delphes 3} detector simulation employs a standard LHC-appropriate parameter card, with jet
clustering performed using the anti-kt algorithm.  The $b$-tagging efficiency is just above 70\% for
transverse momenta between about 65 and 200~GeV, with a mistag rate that climbs from about
a quarter to a third of a percent over this same energy range.
The $\tau$-tagging efficiency is 60\%, with a mistag rate of 1\%.
Selection cuts and computation of collider observables are implemented in the package {\sc AEACuS~3.24}~\cite{aeacus},
and all plots are generated in the companion package {\sc RHADAManTHUS~1.6}~\cite{aeacus}.
While the LHC is currently running at 13 TeV, the larger portion of future integrated statistics are expected to be
collected at an energy of 14 TeV.
Since the processes being studied are of much lower energy than the beam scale, we do not expect our conclusions
to be largely impacted by modifications to the parton distribution and production cross sections between 13 and 14 TeV.
For example, event yield for the $W^+W^-$ background changes by about 10\%.

Following application of the primary topology cuts (Table~\ref{tab:PrimaryCuts}),
we choose higher level cuts which are guided by 1) the desire to increase the signal-to-background ratio, in
order to ensure that any putative excess is robust against systematic uncertainties, and
2) ensuring that one can obtain good statistical significance with a reasonable integrated
luminosity (around $300~{\rm fb}^{-1}$) at the LHC.  We have observed that the former metric often provides a much more
incisive visual guide to the application of cuts in the current context.
Our selections have been guided at each step by iterative analysis of
cut thresholds in $S/(1+B)$, $S/\sqrt{1+B}$, and $S$.

\subsection{Kinematic Reconstruction and Secondary Event Selections}

We will define ``secondary'' event selections as those that are globally beneficial
to all of the targeted mass splittings $\Delta m = 10 - 60$~GeV, but which
go beyond a basic characterization of the topology
and are enforced in the course of a detailed analysis.
A variety of such cuts can be used to reduce the remaining background,
and we will discuss first those involving $Z$-bosons.
At the outset, the process $pp \rightarrow jZZ$, where one $Z$ decays to $\bar \ell \ell$
and the other decays to $\bar \nu \nu$, can be significantly reduced by rejecting events where the
dilepton invariant mass $m_{\ell \ell}$ is close to $m_Z$, as illustrated in the left panel of FIG.~\ref{fig:SHAPE_3}.
The process $pp \rightarrow jZ$, where the $Z$ decays to $\bar \tau \tau$
and the $\tau$s decay leptonically, can be similarly suppressed by a ditau mass cut~\cite{Han:2014aea,Baer:2014kya},
as illustrated in the right panel of FIG.~\ref{fig:SHAPE_3}.

\PlotPairWide{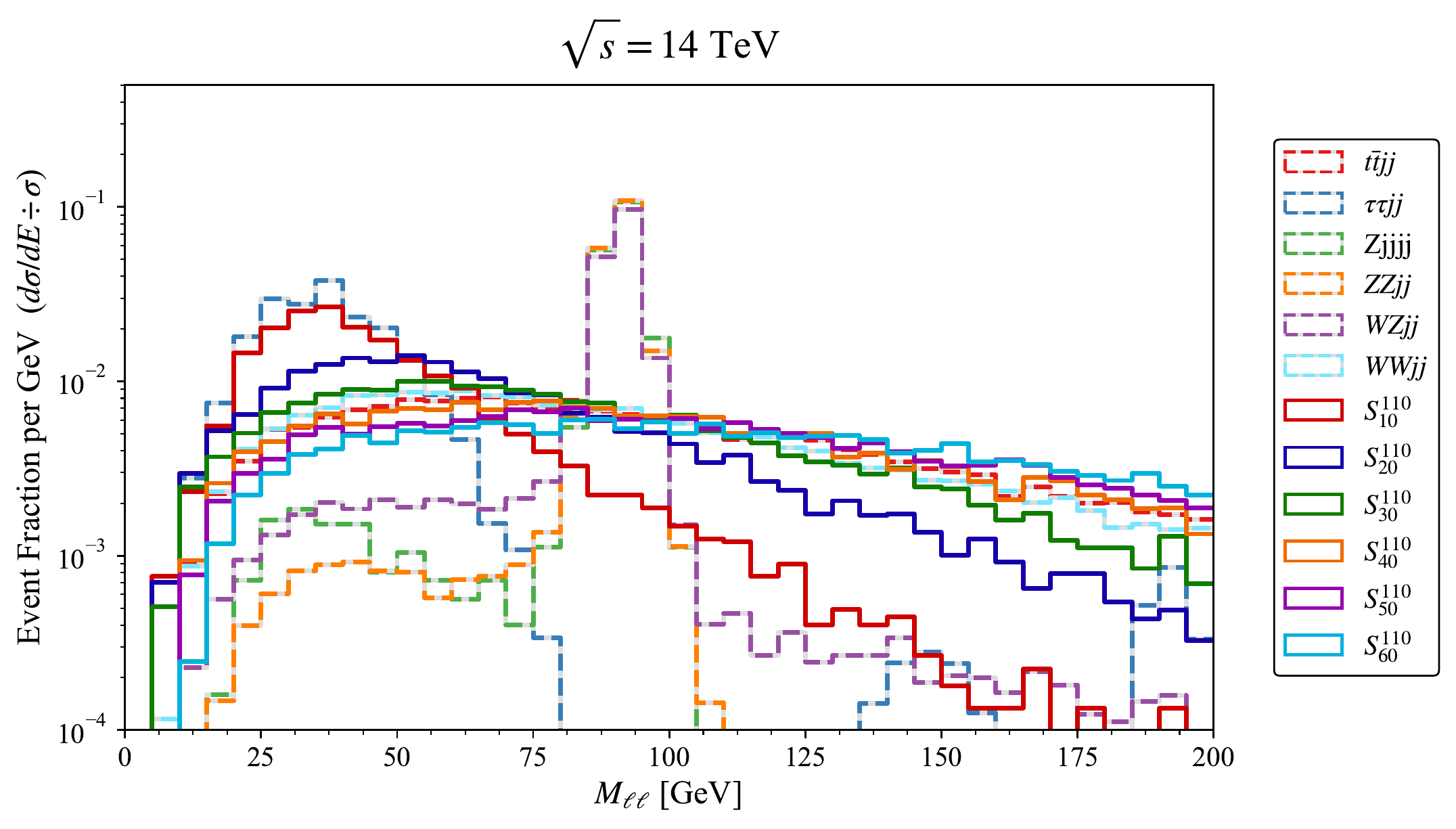}{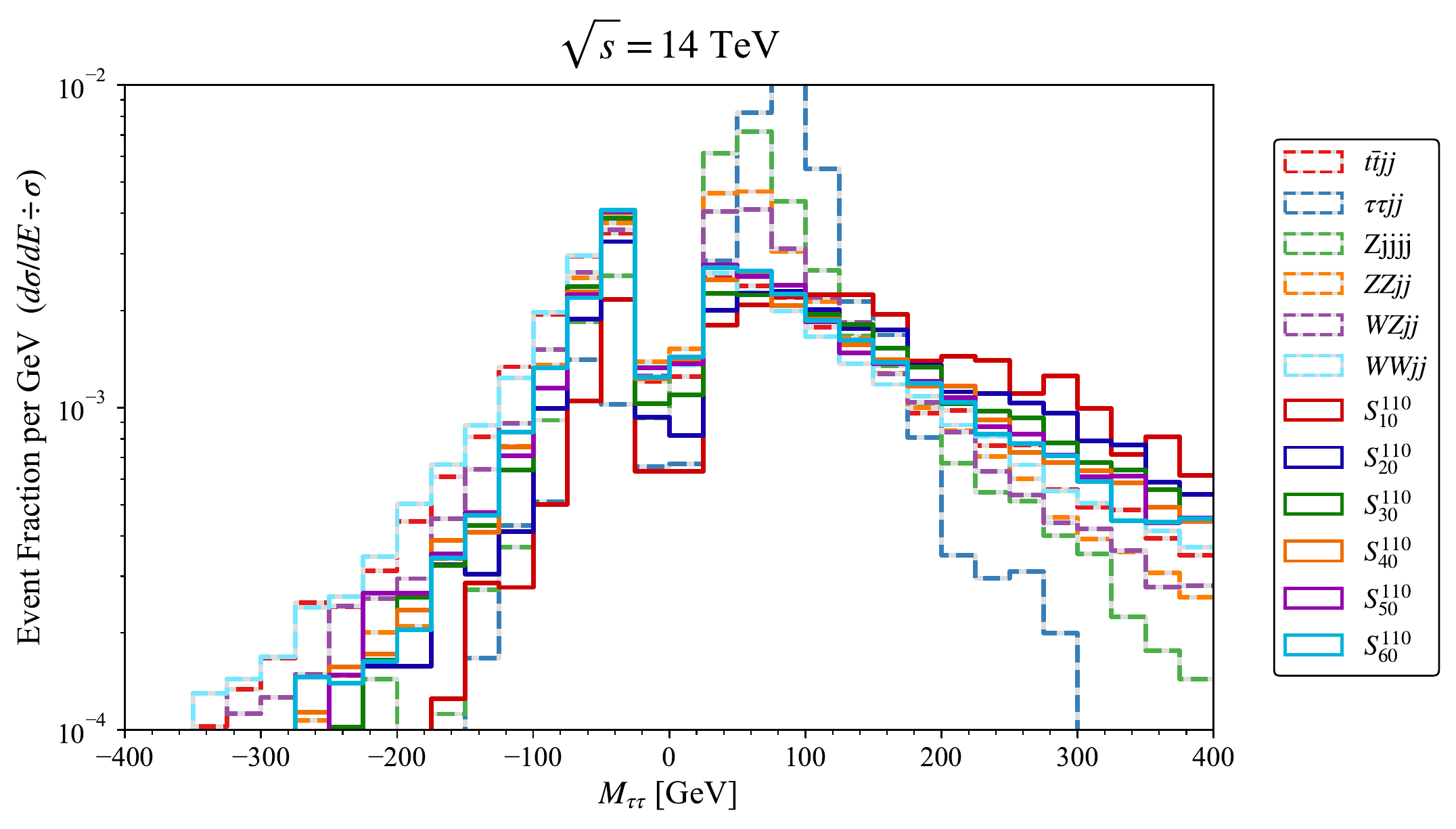}{fig:SHAPE_3}
{
Signal and background event shapes after primary cuts are compared for
(left) the visible dilepton mass $m_{\ell \ell}$, and
(right) the signed ditau mass $m_{\tau \tau} \equiv {\rm sign}\,[m_{\tau \tau}^2]\times\!\sqrt{\vert m_{\tau \tau}^2\vert}$.
The visible dilepton mass reconstruction exhibits a characteristic peak around the $Z$-boson mass for the vector backgrounds.
$m_{\tau\tau}$ is systematically more positive for all signal regions (especially those with narrow mass splitting)
than for the diboson and ditop backgrounds.  Note the ancillary benefits of a lower-bound cut on $m_{\tau\tau}$
to controlling backgrounds beyond just its namesake target.
}

To formulate the ditau mass variable, first proposed in Ref.~\cite{Ellis:1987xu}, one assumes that the entire missing transverse energy of the process arises from two neutrino pairs,
where each pair was emitted collinear with each of the observed leptons and arises (along with the lepton) from the decay of a highly boosted $\tau$.
Momentum conservation in the transverse plane is then sufficient to reconstruct the energy of each neutrino pair, which in turn
determines the momentum of each putative $\tau$ and allows one to reconstruct $m_{\tau \tau}$, the invariant mass of the
putative $\tau$ pair.  In particular, the ditau invariant mass-square may be expressed in closed form as
\bea
m_{\tau \tau}^2 &\equiv& -m_{\ell_1 \ell_2}^2 \frac{(\overrightarrow{P}_T^{\ell_1} \times \overrightarrow{P}_T^{j})
\cdot (\overrightarrow{P}_T^{\ell_2} \times \overrightarrow{P}_T^{j})}{|\overrightarrow{P}_T^{\ell_1} \times \overrightarrow{P}_T^{\ell_2}|^2} ,
\eea
where $m_{\ell_1 \ell_2}^2$ is the invariant squared mass of the lepton system, and $\overrightarrow{P}_T^{\ell_1, \ell_2, j}$ are the transverse
momenta of the leading lepton, subleading lepton, and jet, respectively.  If the leptons and missing momentum arise from the decay of a heavy
particle $X$ with mass $M_X$ to $\bar \tau \tau$, with each boosted $\tau$ decaying leptonically to a collimated system of a lepton and two
neutrinos, then one will find that $m_{\tau \tau}^2 = M_X^2.$\footnote{This is easy to verify.
If $\overrightarrow{P}^{\tau_i} = (1+\zeta_i)\overrightarrow{P}^{\ell_i}$,
where $\zeta_i$ represent the invisible neutrino momentum fraction,
then $M_X^2 = (1+\zeta_1)(1+\zeta_2)m_{\ell_1 \ell_2}^2$.  If $\overrightarrow{P}_X = -\overrightarrow{P}_j = \overrightarrow{P}^{\tau_1}
+\overrightarrow{P}^{\tau_2}$, then
\bea
m_{\tau \tau}^2 &\equiv& -m_{\ell_1 \ell_2}^2 (1+\zeta_1) (1+\zeta_2) \frac{(\overrightarrow{P}_T^{\tau_1} \times \overrightarrow{P}_T^j)
\cdot (\overrightarrow{P}_T^{\tau_2} \times \overrightarrow{P}_T^j)}{|\overrightarrow{P}_T^{\tau_1} \times \overrightarrow{P}_T^{\tau_2}|^2} =
m_{\ell_1 \ell_2}^2 (1+\zeta_1) (1+\zeta_2) ,
\eea
where we have used the fact that, if $\overrightarrow{A}+\overrightarrow{B}=\overrightarrow{C}$, then
$\overrightarrow{A} \times \overrightarrow{C} = \overrightarrow{A} \times \overrightarrow{B} = -\overrightarrow{B} \times \overrightarrow{C}$.
}
The $pp \rightarrow jZ \rightarrow j\bar \tau \tau$  background can thus be removed by a cut on $m_{\tau \tau}$.
Note that $m_{\tau \tau}^2 > 0$ if either $-\overrightarrow{P}_T^{j}$ or $\overrightarrow{P}_T^{j}$ lies between
$\overrightarrow{P}_T^{\ell_1}$ and $\overrightarrow{P}_T^{\ell_2}$,
and is negative if neither do.  This makes it additionally a good kinematic variable for more generally distinguishing event topology,
in addition to rejecting events that literally involve the process $Z \rightarrow \bar \tau \tau \rightarrow \bar \ell \ell +4\nu$.
We should point out here that to use the additional discriminatory power of the variable mentioned above, we used $m_{\tau \tau} \equiv {\rm sign}\,[m_{\tau \tau}^2]\times\!\sqrt{\vert m_{\tau \tau}^2\vert}$ in our analysis as shown in the right panel of FIG.~\ref{fig:SHAPE_3}. This gave our variable a real mass dimension. We checked that the discriminatory power of our redefined variable is the same as the original statistic $m_{\tau \tau}^2$, since $m_{\tau \tau}^2$ is monotonic.

\begin{figure}[!htp]
\centering
\begin{tabular}{ccc}
  \subfloat[$\Delta m = 10$]{\includegraphics[width=55mm]{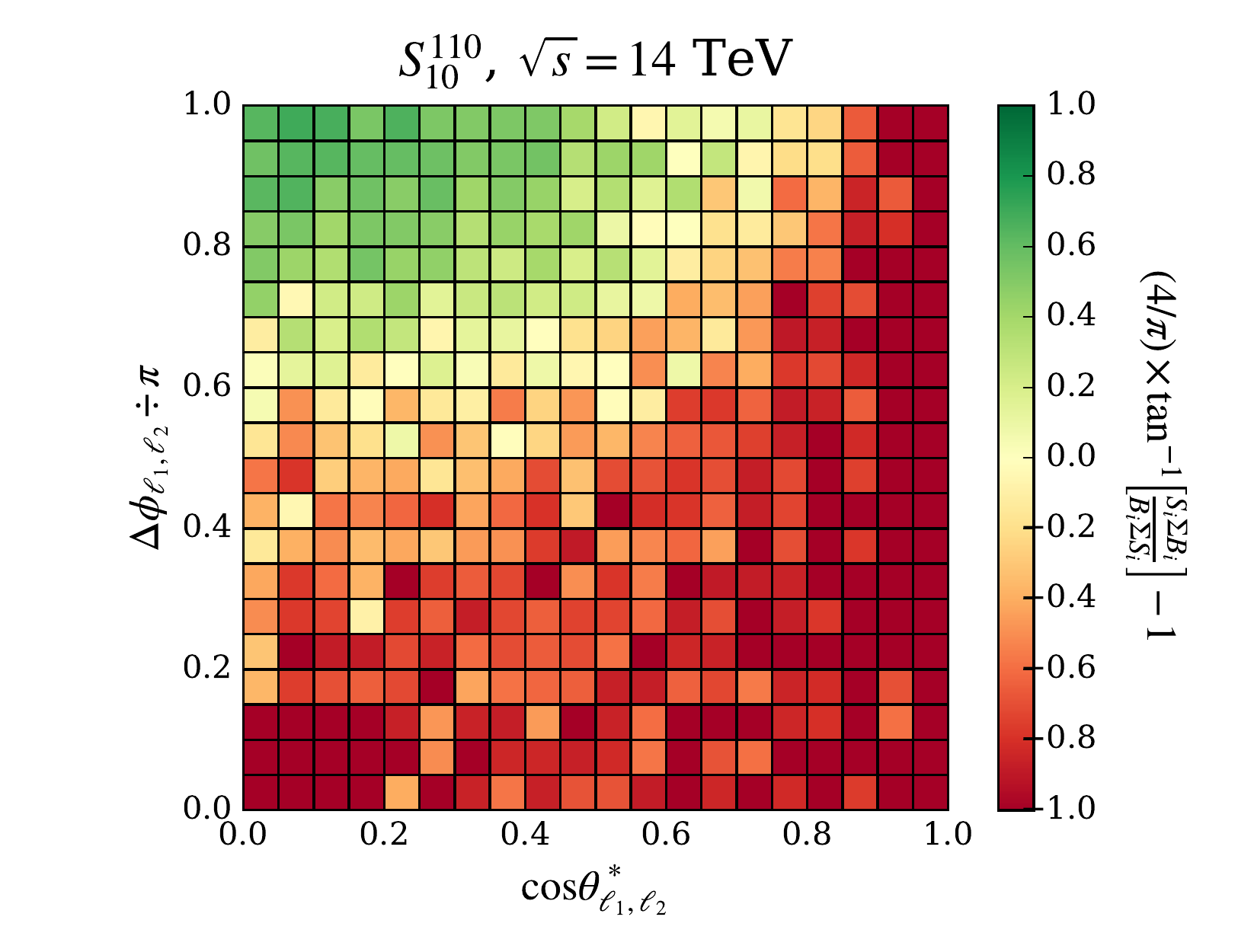}} &
  \subfloat[$\Delta m = 20$]{\includegraphics[width=55mm]{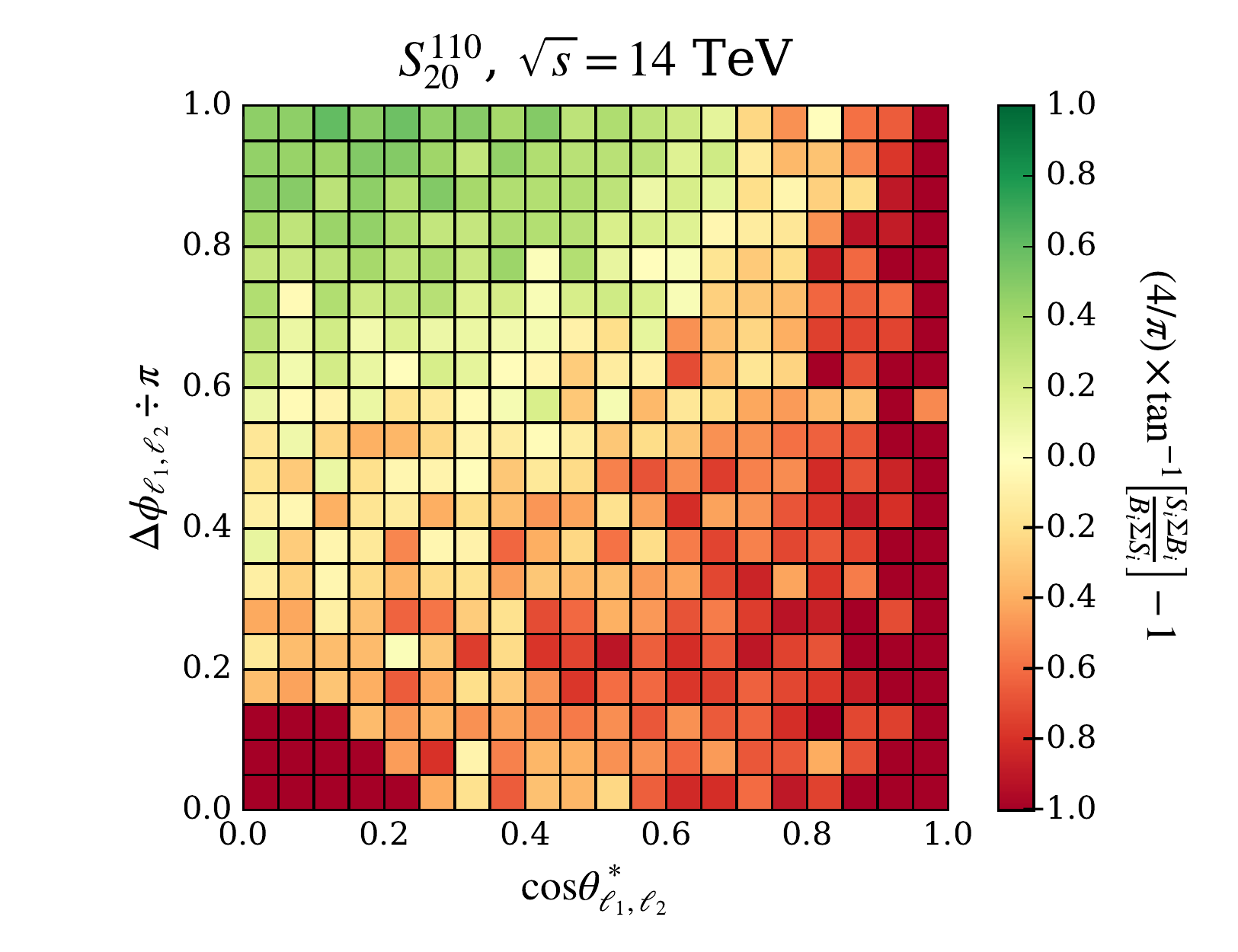}} &
  \subfloat[$\Delta m = 30$]{\includegraphics[width=55mm]{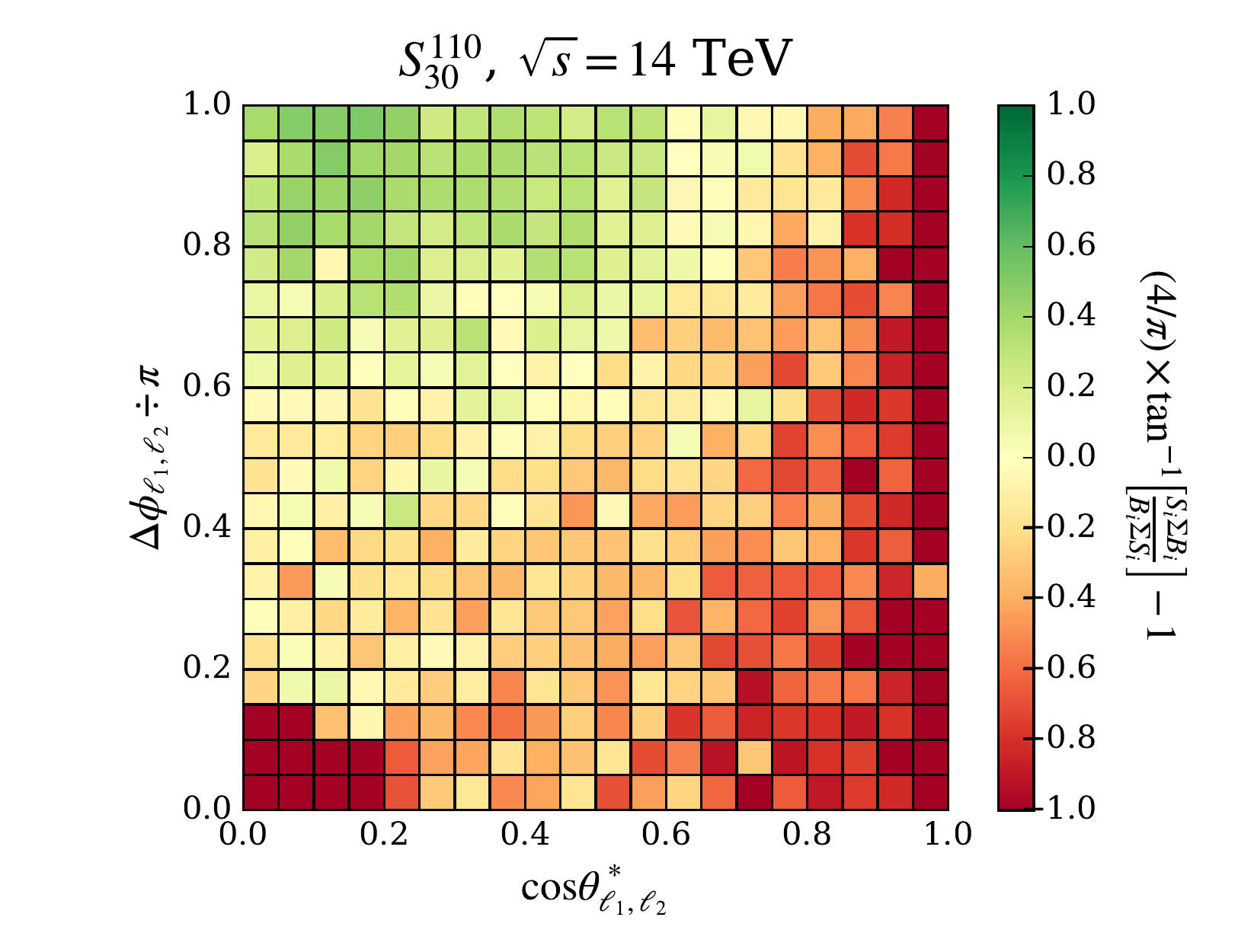}} \\
  \subfloat[$\Delta m = 40$]{\includegraphics[width=55mm]{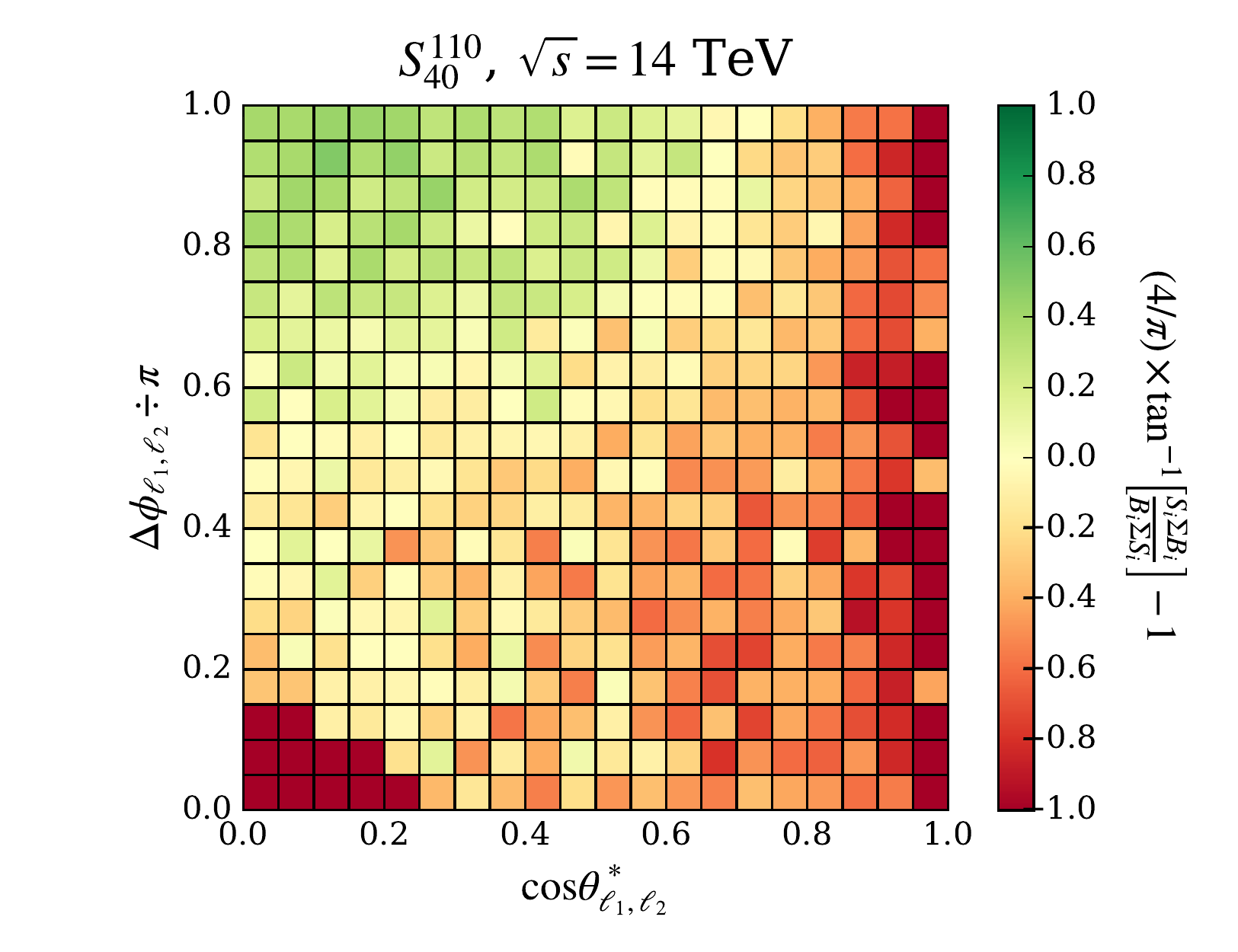}} &
  \subfloat[$\Delta m = 50$]{\includegraphics[width=55mm]{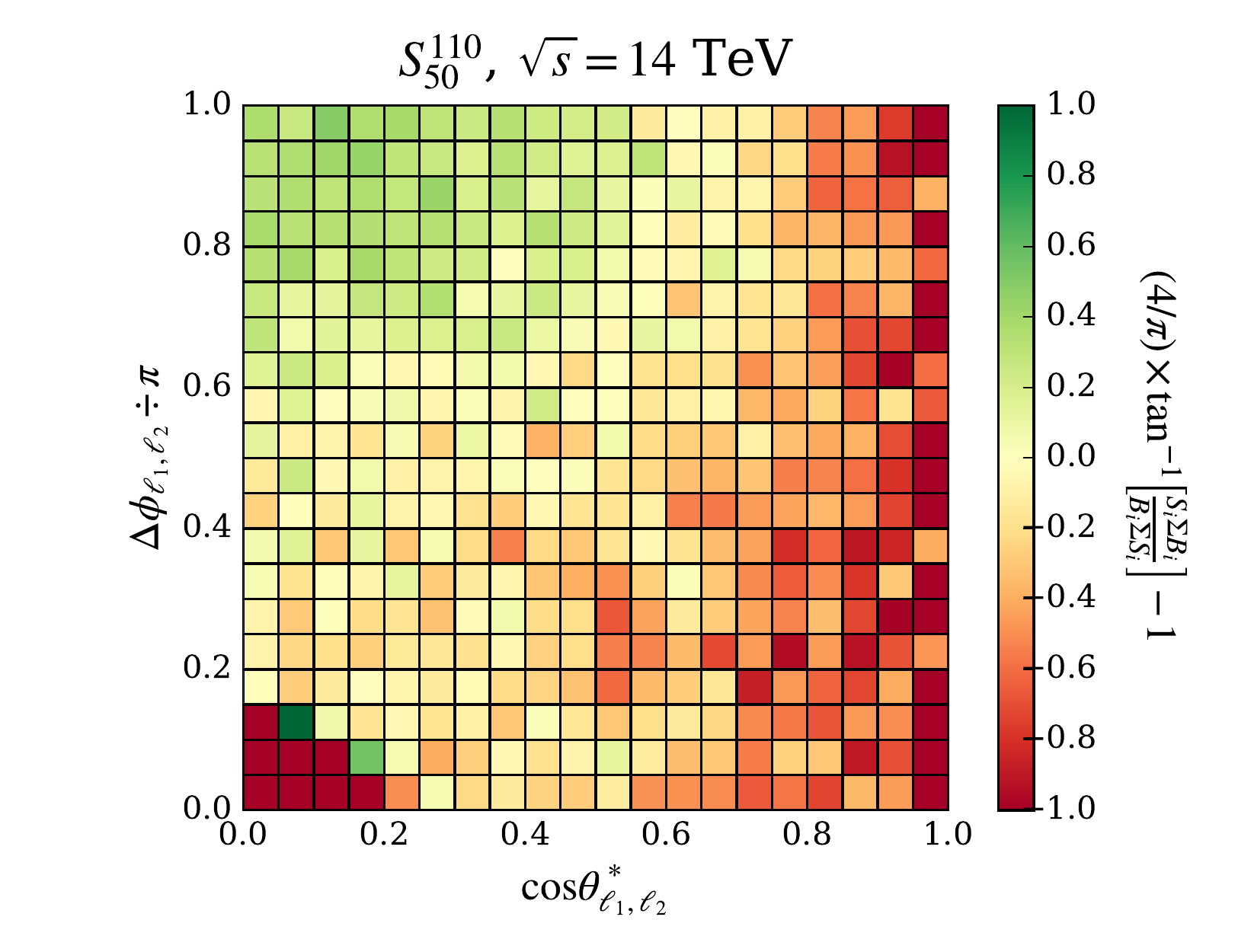}} &
  \subfloat[$\Delta m = 60$]{\includegraphics[width=55mm]{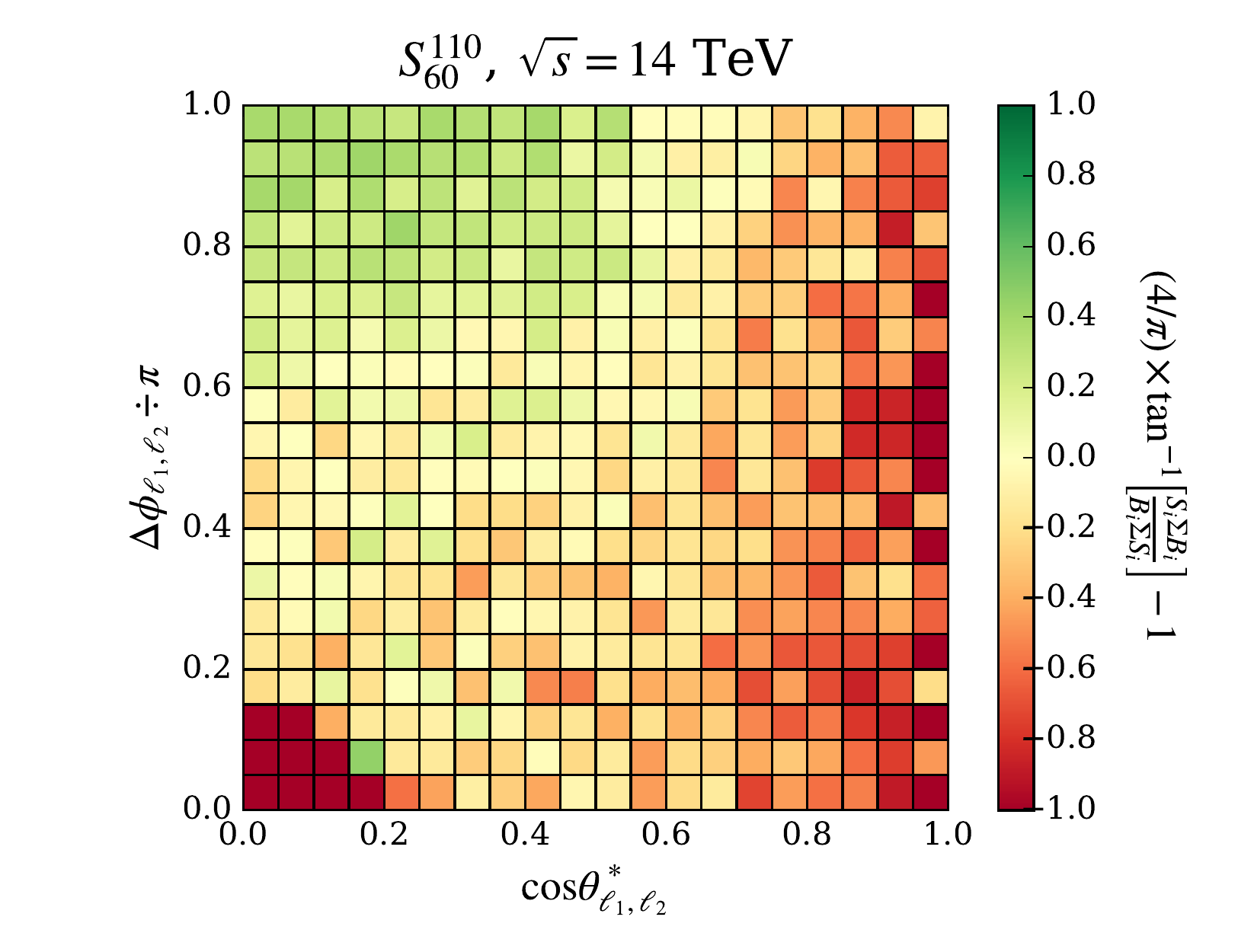}}
\end{tabular}
\begin{minipage}[l]{0.8\linewidth}\vspace{-2pt} \justifying \small {\raggedleft \caption{
A two-dimensional comparison of the variables  $\cos{\theta^\ast}_{\ell_1 \ell_2}$ vs. $\Delta \phi (\ell_1, \ell_2)$
for mass gaps $\Delta m=(10-60)$~GeV
suggests a reasonably consistent region (green) of parameter space in which background is suppressed relative to signal,
although the discrimination power is diffused with increasing mass gap.
The function $(4/\pi) \times \tan^{-1}\left[\frac{S_i \Sigma B_i}{B_i \Sigma S_i}\right]-1$
of the fractional (normalized) cell-wise signal-to-background ratio
is selected as a mapping of $\{0,\infty\} \to \{-1,+1\}$ that is antisymmetric
under the exchange $(S \leftrightarrow B)$.}
\label{fig:SHAPE_4}}\end{minipage}\vspace{0pt}
\end{figure}

One of the most difficult backgrounds to control when searching for the targeted range of models is
the topologically identical $WWj$ process.  Suitable available handles for this discrimination
are differences in the spin of the decaying parent, i.e. spin-0 for the smuon vs. spin-1 for the $W$,
and the mass of the invisible daughter, i.e. up to 100 GeV for the neutralino vs. an essentially massless neutrino.
One tool that is effective for leveraging the first difference is the $\cos \theta^*_{\ell_1 \ell_2}$ variable,
which was introduced in Ref.~\cite{Barr:2005}, and applied in Ref.~\cite{Matchev:2012}
to distinguishing slepton pair-production via Drell-Yan from other processes
that could also yield a lepton pair and missing transverse energy.
It works on the principles that the angular distribution of intermediary
particles with respect to the beam axis in the parton center-of-mass frame
is determined by their spin, and that the lepton angular distribution should reflect this heritage.
Much of the practical utility of the $\cos \theta^*_{\ell_1 \ell_2}$ variable
hinges upon its resiliency against longitudinal boosts of the partonic system.
This feature is apparent in the definition $\cos \theta^*_{\ell_1 \ell_2} \equiv \tanh{(\Delta \eta_{\ell_1 \ell_2}/2)}$, 
where $\Delta \eta_{\ell_1 \ell_2}$ is the pseudorapidity difference between the two leptons, which is 
longitudinal boost-invariant.  
$\cos \theta^*_{\ell_1 \ell_2}$ takes on a
rather simpler geometric interpretation in the
frame where the pseudorapidities of the leptons are equal and opposite,
corresponding there to the cosine of the matched polar angle between each lepton and the beam axis.
The $W$-boson associated backgrounds have a distribution in this variable that is almost flat up to a value around 0.8,
whereas distributions for the scalar-mediated signal models more sharply peak at zero, suggesting
a clear region of preference, as visible in FIGs.~\ref{fig:SHAPE_4}.
Although transverse boosts resulting from the ISR jet and from increasing $\Delta m$
will tend to smear out the described distribution, $\cos \theta^*_{\ell_1 \ell_2}$ remains an effective
tool for distinguishing signal events involving a spin-0 intermediary from background.

Missing transverse energy and jet momentum provide a way of robustly distinguishing signal events from the $\bar t t$ background.
$\bar t t$ events will only survive the primary cuts if both $b$-jets are misidentified and at least one has small
transverse momentum (since the primary cuts reject events with a $b$-jet or with more than one hard jet).
But, in this case, the remaining $b$-jet, as well as the leptons and neutrinos, will tend to have
$p_T \lesssim {\cal O}(m_t)$; such events can be removed by demanding large $\met$ and $P_T^j$.

The optimization of our secondary event selection variables, which are applied equivalently to all signal regions,
suggests first the simple exclusion $m_{\ell\ell} \notin M_Z\pm10$~GeV of dimuon masses within the $Z$-window.
We next apply ${\cos\theta^\ast}_{\ell_1 \ell_2}<0.5$, which discriminates the spin of the states decaying to the leptonic final state.
Subsequently, a uniform cut on the ditau mass $m_{\tau \tau} > 125$~GeV is observed to assist dramatically in the elimination of strong
residual $Z$+Jets backgrounds for all mass regions. We simultaneously elevate the missing transverse energy
and jet momentum cuts to $\met > 125$~GeV and $P_T^{j} > 125$~GeV.
These last three kinematic cuts are observed to  perform well when set to an approximately similar scale.
It is possible in all cases to improve signal-to-background ratio by pushing this trio somewhat harder,
say, up to 175~GeV, at the expense of some statistical significance.

The optimized values of our secondary cuts are summarized in Table~\ref{tab:SecondaryCuts}, 
along with the associated cut flow for signal and background components in terms of the residual cross section.
After the primary cuts, the single vector	
$W$+Jets background (not tabulated) is fully suppressed.
Following the secondary cuts, the $Z$+Jets background is likewise controlled, and the signal-to-background
ratio is roughly (1:3) and (1:2) for the $WWjj$ and $t\bar{t}jj$ components respectively, or
roughly (3:1) and (45:1) for the $WZjj$ and $ZZjj$ components respectively.
The ditau background component has a signal-to-background ratio of approximately (1:1) at this stage of the event selection;
despite the fact that it has not been fully controlled by the $m_{\tau\tau}>125$~GeV cut,
we will show that it does not play a significant role after subsequent the third-level cuts.

\bgroup
\def\arraystretch{1.3}
\begin{table}[!htp]
\caption{Residual effective cross sections (fb) at the LHC14
are tabulated for the $\bar{t}t+$Jets, $\tau\tau+$Jets,  $Z+$Jets, and $VV+$Jets backgrounds, as well
as the six signal benchmarks.  Secondary cuts at this level are applied to all events.\\}
\label{tab:SecondaryCuts}
\scalebox{1.02}{%
\begin{tiny}
\hspace{-46pt}
\begin{tabular}{|c||c|c|c|c|c|c||c|c||c|c||c|c|}
\hline
Selection & $t\bar{t}jj$ & $\tau\tau jj$ & $Zjjjj$ & $ZZjj$ & $WZjj$ & $WWjj$ & $S^{110}_{10}$ & $S^{110}_{20}$ & $S^{110}_{30}$ & $S^{110}_{40}$ & $S^{110}_{50}$ & $S^{110}_{60}$ \\
\hline
\hline
$m_{\ell\ell} \notin M_Z\pm10$~GeV & $1.4 \times 10^{2}$ & $1.8 \times 10^{2}$ & $6.2 \times 10^{2}$ & $2.0 \times 10^{0}$ & $1.0 \times 10^{1}$ & $7.9 \times 10^{1}$ & $6.0 \times 10^{0}$ & $9.2 \times 10^{0}$ & $1.1 \times 10^{1}$ & $1.2 \times 10^{1}$ & $1.3 \times 10^{1}$ & $1.4 \times 10^{1}$ \\
\hline
$\cos \theta^*_{ \ell_1,\ell_2} < 0.5$ & $8.1 \times 10^{1}$ & $1.6 \times 10^{2}$ & $4.7 \times 10^{2}$ & $1.4 \times 10^{0}$ & $6.7 \times 10^{0}$ & $4.5 \times 10^{1}$ & $4.8 \times 10^{0}$ & $6.9 \times 10^{0}$ & $8.0 \times 10^{0}$ & $9.0 \times 10^{0}$ & $9.5 \times 10^{0}$ & $1.0 \times 10^{1}$ \\
\hline
$m_{\tau\tau} > 125$~GeV & $2.7 \times 10^{1}$ & $2.3 \times 10^{1}$ & $8.7 \times 10^{1}$ & $3.0 \times 10^{-1}$ & $1.4 \times 10^{0}$ & $1.4 \times 10^{1}$ & $3.0 \times 10^{0}$ & $3.4 \times 10^{0}$ & $3.6 \times 10^{0}$ & $3.9 \times 10^{0}$ & $4.1 \times 10^{0}$ & $4.3 \times 10^{0}$ \\
\hline
$\met > 125$~GeV & $2.9 \times 10^{0}$ & $6.6 \times 10^{-1}$ & 0 & $1.5 \times 10^{-2}$ & $2.2 \times 10^{-1}$ & $2.3 \times 10^{0}$ & $5.1 \times 10^{-1}$ & $5.8 \times 10^{-1}$ & $6.6 \times 10^{-1}$ & $7.1 \times 10^{-1}$ & $7.9 \times 10^{-1}$ & $8.9 \times 10^{-1}$ \\
\hline
Jet $P_T>125$~GeV & $1.1 \times 10^{0}$ & $6.6 \times 10^{-1}$ & 0 & $1.1 \times 10^{-2}$ & $1.9 \times 10^{-1}$ & $1.7 \times 10^{0}$ & $4.9 \times 10^{-1}$ & $5.2 \times 10^{-1}$ & $5.2 \times 10^{-1}$ & $4.6 \times 10^{-1}$ & $4.5 \times 10^{-1}$ & $4.5 \times 10^{-1}$ \\
\hline
\end{tabular}
\end{tiny}}
\end{table}
\egroup

\subsection{Angular Distributions and Tertiary Event Selections}

We will define ``tertiary'' event selections as those whose impact is
differentially correlated with the specific value of the mass splitting $\Delta m$.
The remaining backgrounds at this stage of the flow are dominated by $t\bar{t}(j)$, $\tau\tau j$, and $WWj$ production.
To further distinguish signal events from SM background, one must contrast the energy and angular distribution
of the leptons and invisible particles arising from signal events with those arising from background events.
Several types of kinematic variables are useful here:
\begin{itemize}
\item{The relative strengths of the missing transverse energy $\met$ and the jet momentum $P_T^j$. }
\item{The angle $\Delta \phi(\met,j)$ between the leading jet and the
missing transverse energy. This variable and that prior are very useful to identifying the small mass gap scenarios. See FIG.~\ref{fig:SHAPE_2}.}
\item{The angle $\Delta \phi (\met, \ell_{1,2})$ between either lepton and the
missing transverse energy.  These variables measure the collimation of the lepton plus invisible system, and demanding
smaller values of this angle favors slepton signal events over $\bar t t (j)$ background events.
See FIG.~\ref{fig:SHAPE_2}.}
\item{$P_T^{\ell_2}$, the transverse momentum of the sub-leading lepton.  This variable
measures the hardness of the lepton plus invisible system; demanding a larger value benefits processes with larger mass gaps.
Note that a lower bound on this variable implicitly includes a similar bound on the leading lepton. See FIG.~\ref{fig:SHAPE_1}.}
\item{The angle $\Delta \phi (\ell_1,\ell_2)$ between the leading lepton and the subleading lepton,
which is related to the topology and inherited boost of decays into the leptonic system.  Demanding larger values of
this angle favors processes with heavier parents. See FIGs.~\ref{fig:SHAPE_4},~\ref{fig:SHAPE_1}.}
\end{itemize}

\PlotPairWide{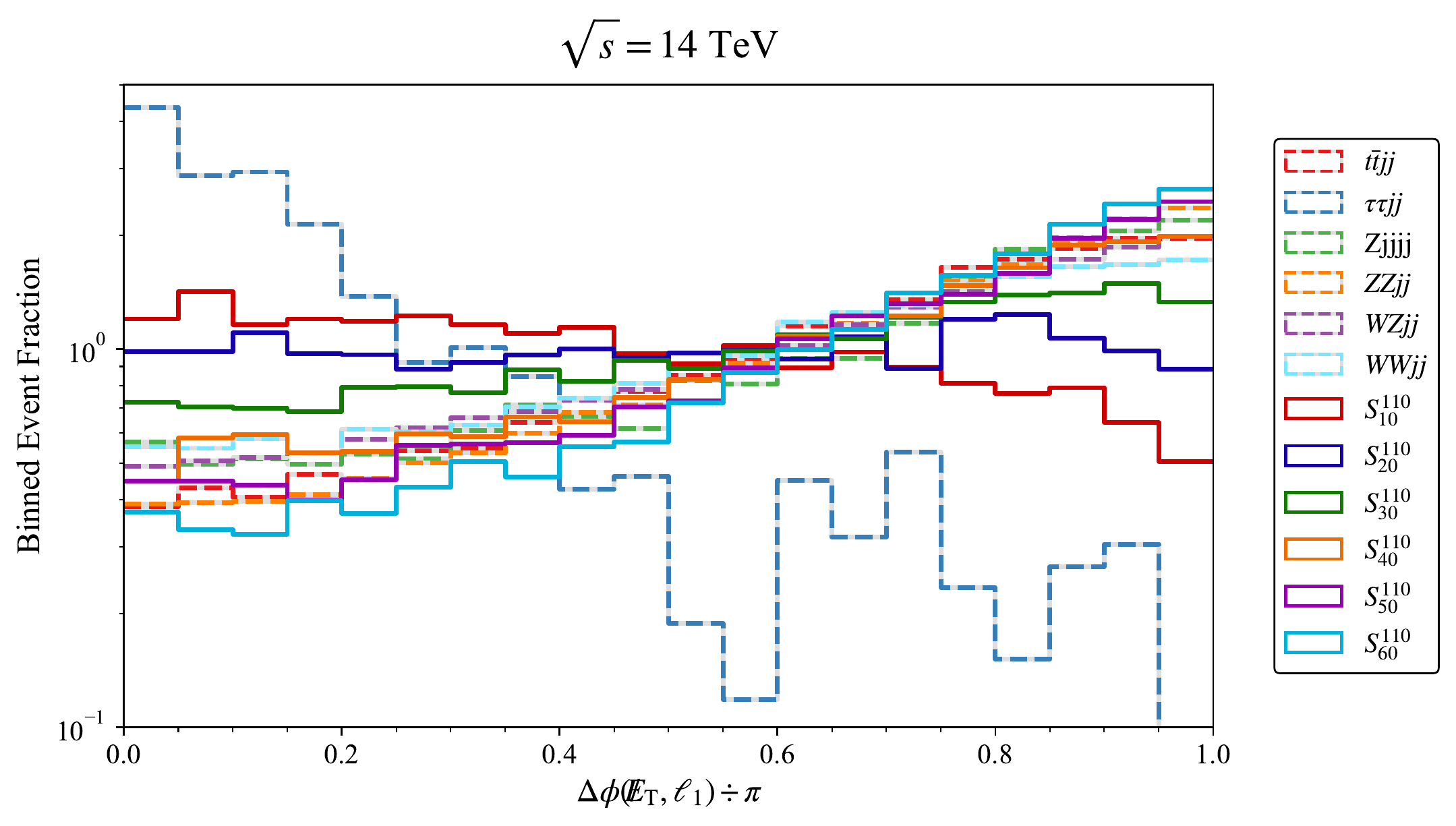}{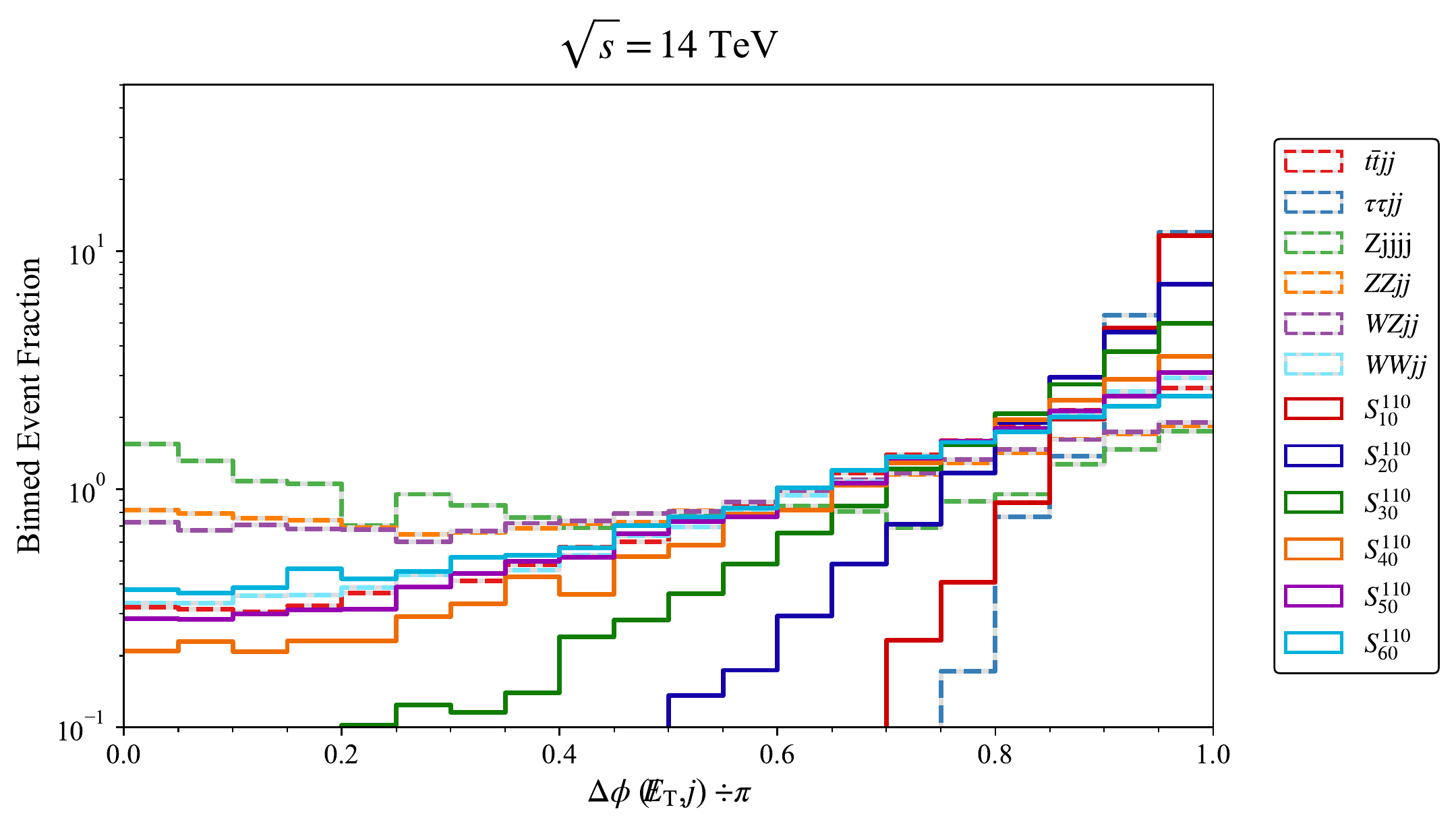}{fig:SHAPE_2}
{
Signal and background event shapes after primary cuts (discussed in Table~\ref{tab:PrimaryCuts}) are compared for
the azimuthal angular separation between the missing transverse energy and
(left) the leading jet $\Delta \phi (\met,j)$ or (right) the leading lepton $\Delta \phi (\met,\ell_1)$ .
The leading lepton is systematically more aligned with the $\met$ for signal regions with narrow mass splitting than for background.
This statistic can behave favorably as an upper bound for the lighter mass gaps and as a lower bound for the heavier mass gaps.
The leading jet is systematically more anti-aligned with the $\met$ for signal regions with narrow mass splitting than for background.
}

\PlotPairWide{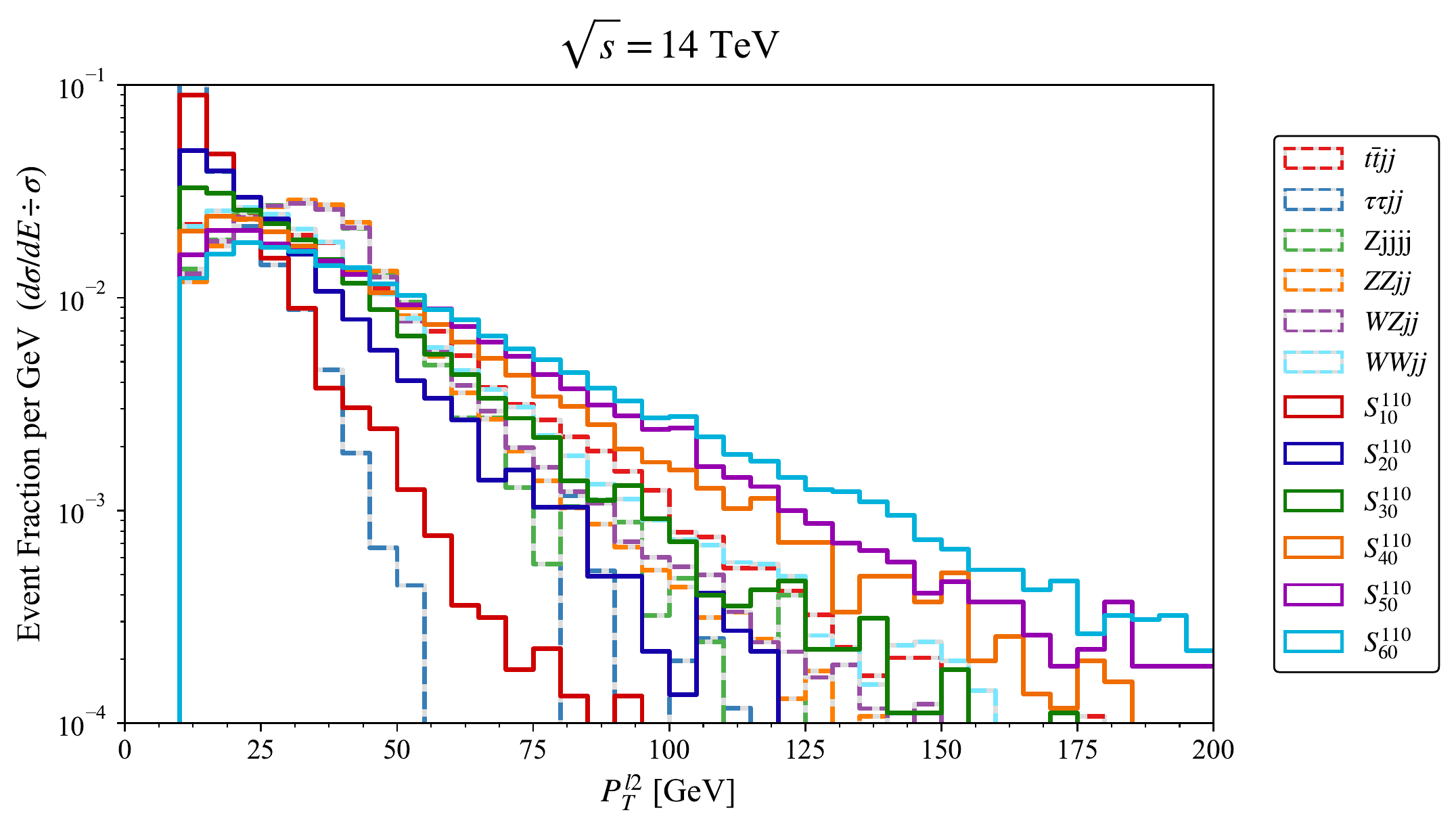}{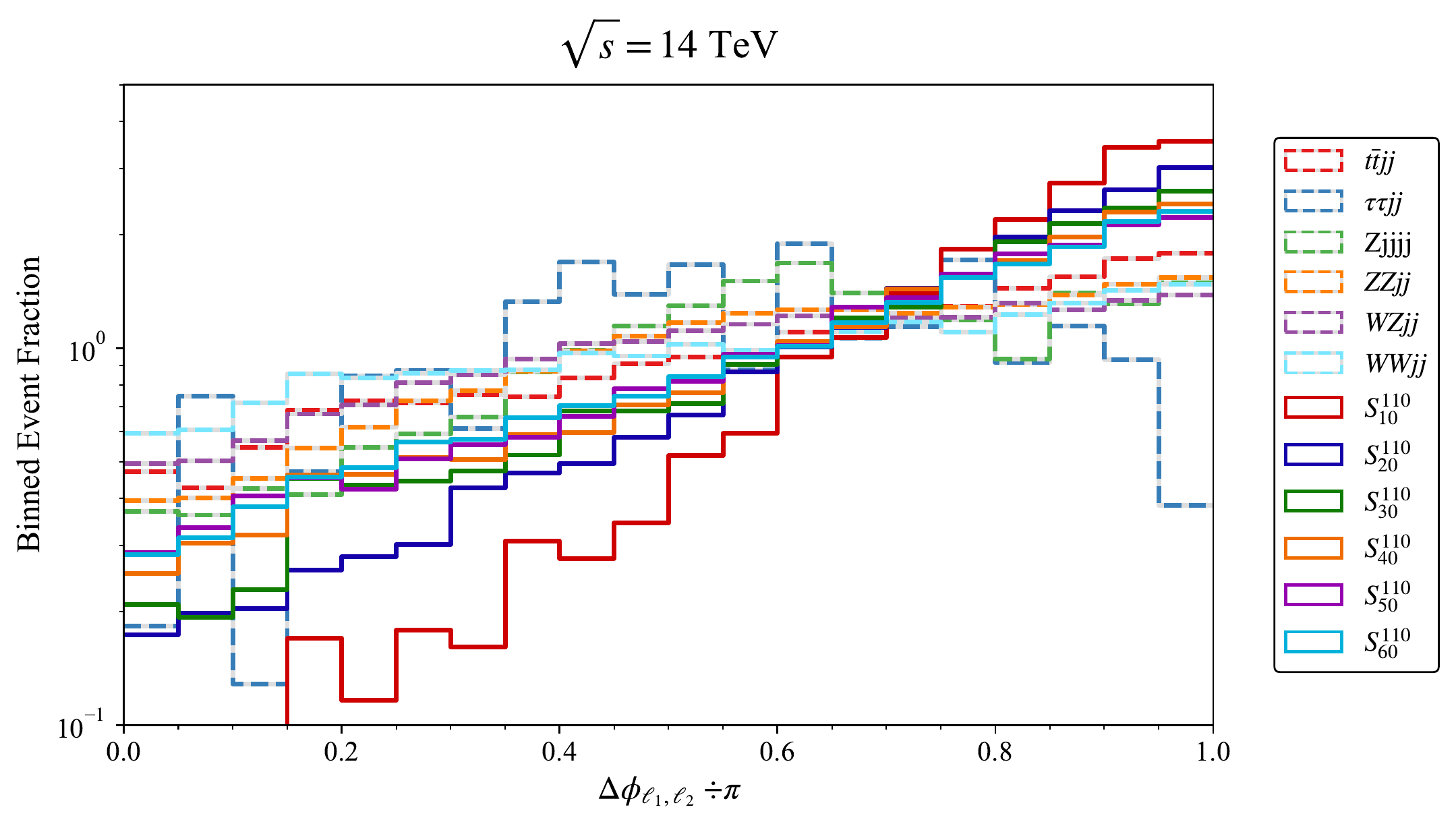}{fig:SHAPE_1}
{
Signal and background event shapes after primary cuts (discussed in Table~\ref{tab:PrimaryCuts}) are compared for
(left) the minimal leptonic transverse momentum $P_T^{\ell_2}$ and
(right) the azimuthal angular separation between the visible dilepton pair $\Delta \phi (\ell_1, \ell_2)$.
The value of $P_T^{\ell_2}$ for models with a mass gap $\Delta m \simeq 30$~GeV closely mimics
the background from SM vector decays, whereas smaller mass gaps lead to a softer leptonic system, and vice versa.
The dilepton signal pairs are all systematically more widely separated
than those arising from SM backgrounds, although the effect becomes more pronounced
when the mass splitting is small.
}

To remove the $jWW$ background, one can utilize the angular distribution
of the leptons.
For this purpose, it is convenient to consider the center-of-mass frame for the system consisting of all
particles except the hard jet (the unboosted system).
For this system, both signal and
background processes consist of the production of two intermediary particles ($\tilde \ell^* \tilde \ell$ and $W^+W^-$, respectively), followed
by the decay of each intermediary to a lepton and an invisible particle ($\chi$ and $\nu$, respectively).
In both cases, the leptons will tend to be anti-collimated; if the
intermediaries are produced at threshold then the lepton distribution will tend to be isotropic, but if the intermediaries are produced with reasonable longitudinal boost (not to be confused with transverse boost coming from hard jets) then their decay products will tend to be collimated with the direction of the parent, yielding leptons which are anti-collimated.
The net result will be that the leptons produced from both $\tilde \ell^* \tilde \ell$ and $W^+W^-$ production processes will be biased towards
anti-collimation, in the center-of-mass frame of the unboosted system.

The transverse boost resulting from emission of a jet will tend to smear out these angular distributions.
But the heavier the intermediary, the less the distribution will
be smeared, since heavier intermediaries acquire a smaller boost from the emission of a jet with fixed momentum.  As a result, provided the
slepton is heavier than $m_W$, the leptons arising from $\tilde \ell^* \tilde \ell j$ production will tend to be more anti-collimated, in
the frame of the detector, than the leptons arising from $WWj$ backgrounds.  Moreover, one expects this bias to increase as the sleptons are made
heavier.

Conversely, the leading lepton produced from $\tilde \ell^* \tilde \ell j$ production will tend to be collimated with the missing transverse
energy, since both arise from the decay products of a boosted slepton.  This correlation can be used to distinguish signal from the
$\bar t t (j)$ background.  The angle between the leading lepton and the missing transverse energy is much more uniform in $\bar t t$ background
events, since a significant portion of the $\met$ in these events arises from mismeasured or missed jets.

Finally, the relationship between $\met$ and $P_T^j$ is useful in distinguishing the scenario in which the
bino-slepton mass gap is small.  In this scenario, slepton decay produces leptons and binos with very small
momentum in the unboosted frame.  The boosted slepton system has a momentum equal and opposite to that of the
jet, but most of this momentum is carried by the heavier binos.  As a result, signal event in the nearly-degenerate
regime tend to have $P_T^j / \met \gtrsim 1$, with $\met$ and $P_T^j$ largely anti-collimated.

Note that one generally expects it to be difficult to trust a determination of missing transverse energy which is
either parallel or anti-parallel to a jet, since jet mismeasurement can cause an apparent missing
transverse energy  along the axis of a jet.  But we will always require at least $\met \gtrsim 125\gev$,
and jet mismeasurement is generally a less significant source of uncertainty when the invisible system is harder.

\section{Final Optimization and Results}

In the prior subsection we outlined a set of general considerations governing the selection of tertiary cuts which can distinguish
signal from background in the case of a bino-slepton pair with a squeezed spectrum.
In this section we provide a detailed application of these cuts differentially to each 
benchmark mass-splitting ans\"atz, and subsequently describe our main results.

We remark at this stage that many of the kinematic discriminants considered in the present study
exhibit strong correlations, such that multiple semi-equivalent solutions to the optimization puzzle
exist in certain cases.  Relatedly, the relevant signal vs.~background distribution
shapes may be altered significantly for the events residual at a certain stage of the flow,
as one proceeds successively through the application of cuts.  Certain selections which do not
visually present a strong discrimination in FIGs.~\ref{fig:SHAPE_3}--\ref{fig:SHAPE_2} at the
stage of primary cuts may nevertheless prove to appreciably elevate
the signal-to-background ratio, for example, after more cuts have been applied.  Finally, given that
a certain fraction of the topologically matched $WWjj$ background is (apparently irreducibly)
kinematically similar to the signal profile, it becomes beneficial to
select relatively soft cuts in certain variables where the signal-to-background
resolving power is marginal, in order to emphasize statistical resolution.
One might alternatively consider the use of Bayesian optimization techniques,
as recently outlined in Ref.~\cite{Alves:2017ued},
for systematic maximization of both the statistical significance and $S/B$.

The identities and values of these tertiary cuts are detailed in Table~\ref{tab:tertiary},
along with the cross section flow of key event populations,
for three scenarios tailored respectively to the narrower, intermediate, and wider mass gap regimes.
The intermediate selection scenario provides reasonably balanced performance for all values of $\Delta m$,
although it is possible to do markedly better for small and large values by a more specialized approach.
Also tabulated are the estimated number of signal events (with $300~{\rm fb}^{-1}$
integrated luminosity), signal-to-background ratios, and signal significances after the application of all cuts.
We describe each of these three optimizations briefly in the following subsections.

\bgroup
\def\arraystretch{1.3}
\begin{table}[!htp]
\caption{Residual effective cross sections (fb) after the application of tertiary cuts targeted at smaller, intermediate, and larger mass gaps.
Also given are the number of signal events, $S/(1+B)$, and projected significance at $300~{\rm fb}^{-1}$.  For each tertiary cut group, the
models in the targeted mass gap range are in bold-face.}
\label{tab:tertiary}
\begin{center}
\scalebox{1.0}{%
\begin{tiny}
\hspace{-45pt}
\begin{tabular}{|c||c|c|c|c|c||c|c||c|c||c|c|}
\hline
Selection & $t\bar{t}jj$ & $\tau\tau jj$ & $ZZjj$ & $WZjj$ & $WWjj$ & $S^{110}_{10}$ & $S^{110}_{20}$ & $S^{110}_{30}$ & $S^{110}_{40}$ & $S^{110}_{50}$ & $S^{110}_{60}$ \\
\hline
\hline
\multicolumn{12}{|c|}{Small Mass Gap Optimization} \\
\hline
$1.0 < P_T^j \div \met < 1.3$ & $4.5 \times 10^{-1}$ & $5.5 \times 10^{-3}$ & $2.7 \times 10^{-3}$ & $6.2 \times 10^{-2}$ & $6.6 \times 10^{-1}$ & $\bm{4.2 \times 10^{-1}}$ & $\bm{3.0 \times 10^{-1}}$ & $2.4 \times 10^{-1}$ & $1.8 \times 10^{-1}$ & $1.6 \times 10^{-1}$ & $1.4 \times 10^{-1}$ \\
\hline
$\Delta \phi(\,\met,j)\div \pi > 0.95$ & $1.8 \times 10^{-1}$ & $5.5 \times 10^{-3}$ & $2.2 \times 10^{-3}$ & $2.7 \times 10^{-2}$ & $3.8 \times 10^{-1}$ & $\bm{4.0 \times 10^{-1}}$ & $\bm{2.3 \times 10^{-1}}$ & $1.7 \times 10^{-1}$ & $9.6 \times 10^{-2}$ & $7.7 \times 10^{-2}$ & $6.2 \times 10^{-2}$ \\
\hline
Events at $\mathcal{L}=300~{\rm fb}^{-1}$ & 52.7 & 1.7 & 0.7 & 8.1 & 113.6 & {\bf 120.0} & {\bf 69.0} & 51.0 & 28.8 & 23.1 & 18.6\\
\hline
$S \div (1+B)$ & - & - & - & - & -& {\bf 0.68} & {\bf 0.39} & 0.29 & 0.16 & 0.13 & 0.10\\
\hline
$S \div \sqrt{1+B}$ & - & - & - & - & -&{\bf 9.0} & {\bf 5.2} & 3.8 & 2.2 & 1.7 & 1.4\\
\hline
\hline
\multicolumn{12}{|c|}{Intermediate Mass Gap Optimization} \\
\hline
$\Delta \phi(\ell_1,\ell_2)\div \pi > 0.5$ & $1.1 \times 10^{0}$ & $5.5 \times 10^{-3}$ & $7.7 \times 10^{-3}$ & $1.2 \times 10^{-1}$ & $1.3 \times 10^{0}$ & $4.0 \times 10^{-1}$ & $4.0 \times 10^{-1}$ & $\bm{4.4 \times 10^{-1}}$ & $\bm{4.1 \times 10^{-1}}$ & $3.7 \times 10^{-1}$ & $3.9 \times 10^{-1}$ \\
\hline
$\Delta \phi(\,\met,\ell_1)\div \pi < 0.6$ & $4.8 \times 10^{-1}$ & $5.5 \times 10^{-3}$ & $5.5 \times 10^{-3}$ & $7.9 \times 10^{-2}$ & $9.0 \times 10^{-1}$ & $3.7 \times 10^{-1}$ & $3.3 \times 10^{-1}$ & $\bm{3.3 \times 10^{-1}}$ & $\bm{3.0 \times 10^{-1}}$ & $2.4 \times 10^{-1}$ & $2.1 \times 10^{-1}$ \\
\hline
$\Delta \phi(\,\met,\ell_2)\div \pi < 0.6$ & $1.8 \times 10^{-1}$ & 0.0 & $4.4 \times 10^{-3}$ & $4.8 \times 10^{-2}$ & $5.1 \times 10^{-1}$ & $2.7 \times 10^{-1}$ & $2.3 \times 10^{-1}$ & $\bm{2.2 \times 10^{-1}}$ & $\bm{2.0 \times 10^{-1}}$ & $1.6 \times 10^{-1}$ & $1.4 \times 10^{-1}$ \\
\hline
Events at $\mathcal{L}=300~{\rm fb}^{-1}$ & 52.8 & 0.0 & 1.3 & 14.5 & 151.7 & 81.0 & 69.0 & {\bf 66.0} & {\bf 60.0} & 48.0 & 42.0\\
\hline
$S \div (1+B)$ & - & - & - & - & - & 0.37 & 0.31 & {\bf 0.30} & {\bf 0.27} & 0.22 & 0.19\\
\hline
$S \div \sqrt{1+B}$ & - & - & - & - & - & 5.4 & 4.6 & {\bf 4.4} & {\bf 4.0} & 3.2 & 2.8\\
\hline
\hline
\multicolumn{12}{|c|}{Large Mass Gap Optimization} \\
\hline
$\Delta \phi(\,\met,\ell_1)\div \pi > 0.25$ & $8.5 \times 10^{-1}$ & 0.0 & $6.6 \times 10^{-3}$ & $1.0 \times 10^{-1}$ & $9.5 \times 10^{-1}$ & $2.4 \times 10^{-1}$ & $2.7 \times 10^{-1}$ & $3.3 \times 10^{-1}$ & $2.9 \times 10^{-1}$ & $\bm{3.0 \times 10^{-1}}$ & $\bm{3.3 \times 10^{-1}}$ \\
\hline
$P_T^{\ell_2} > 40$~GeV & $3.4 \times 10^{-1}$ & 0.0 & $5.6 \times 10^{-4}$ & $3.7 \times 10^{-2}$ & $4.1 \times 10^{-1}$ & $1.1 \times 10^{-2}$ & $7.3 \times 10^{-2}$ & $1.4 \times 10^{-1}$ & $1.5 \times 10^{-1}$ & $\bm{2.1 \times 10^{-1}}$ & $\bm{2.4 \times 10^{-1}}$ \\
\hline
Events at $\mathcal{L}=300~{\rm fb}^{-1}$ & 102.2 & 0.0 & 0.2 & 11.0 & 124.3 & 3.3 & 21.9 & 42.0 & 45.0 & {\bf 63.0} & {\bf 72.0}\\
\hline
$S \div (1+B)$ & - & - & - & - & - & 0.01 & 0.09 & 0.18 & 0.19 & {\bf 0.26} & {\bf 0.30}\\
\hline
$S \div \sqrt{1+B}$ & - & - & - & - & - & 0.2 & 1.4 & 2.7 & 2.9 & {\bf 4.1} & {\bf 4.7}\\
\hline
\end{tabular}
\end{tiny}}
\end{center}
\end{table}
\egroup

\subsection{$\Delta m=10,\,20\gev$}
If $\Delta m$ is small, $\sim 10\gev$, then one finds that the signal lepton plus invisible system tends to be
relatively soft, and collimated anti-parallel to the jet momentum, as compared to background events.
We emphasize here interesting alternatives to selecting directly for low $p_T^{\ell_{1,2}}$~\cite{Han:2014aea}
(although the presented selections are certainly correlated with the presence of soft leptons).
As argued previously, the softness of the leptons is indeed correlated with a ratio $P_T^j / \met \sim 1$, as well
as with the anti-collimation of $\met$ and $P_T^j$.
In particular, we enforce a bound $1<p^j_T/\met<1.3$ on the jet and missing transverse momenta,
in conjunction with a requirement that the $\met$ and jet are also back-to-back, i.e.
$\Delta \phi(\,\met,j)\sim \pi$, within about 5\% (see FIG.~\ref{fig:SHAPE_1}).
We show the cut flows in Table \ref{tab:tertiary}. The significance can be above 5-9~$\sigma$
for a luminosity of $300~{\rm fb}^{-1}$ with $S/B$ around 40-70\%.  Note that the signal-to-background
ratio found through this analysis is significantly larger than that found using the selection cuts of~~\cite{Han:2014aea}.

\subsection{$\Delta m=30,\,40\gev$}
Angular correlation cuts are very useful in this mass gap range to remove the $W,\,Z$ backgrounds.
We turn first to the azimuthal separation $\Delta \phi (\ell_1, \ell_2)$ between
the visible leptons.
The cut $\Delta \phi (\ell_1, \ell_2)/\pi > 0.5$ is applied first, as motivated
in the right panel of FIG.~\ref{fig:SHAPE_1}, and also FIGs.~\ref{fig:SHAPE_4}.
Next, an upper bound $\Delta \phi (\ell_{1,2}, \met)/ \pi < 0.6$ on the angle between the
leptons and the missing transverse momentum vector completes the selection, and substantially
elevates the signal-to-background ratio, as shown in the left panel of FIG.~\ref{fig:tertiary}.
Overall, the significance after tertiary cuts can be around 4 for luminosity $300~{\rm fb}^{-1}$ and $S/B$ is around 30\%.

\PlotPairWide{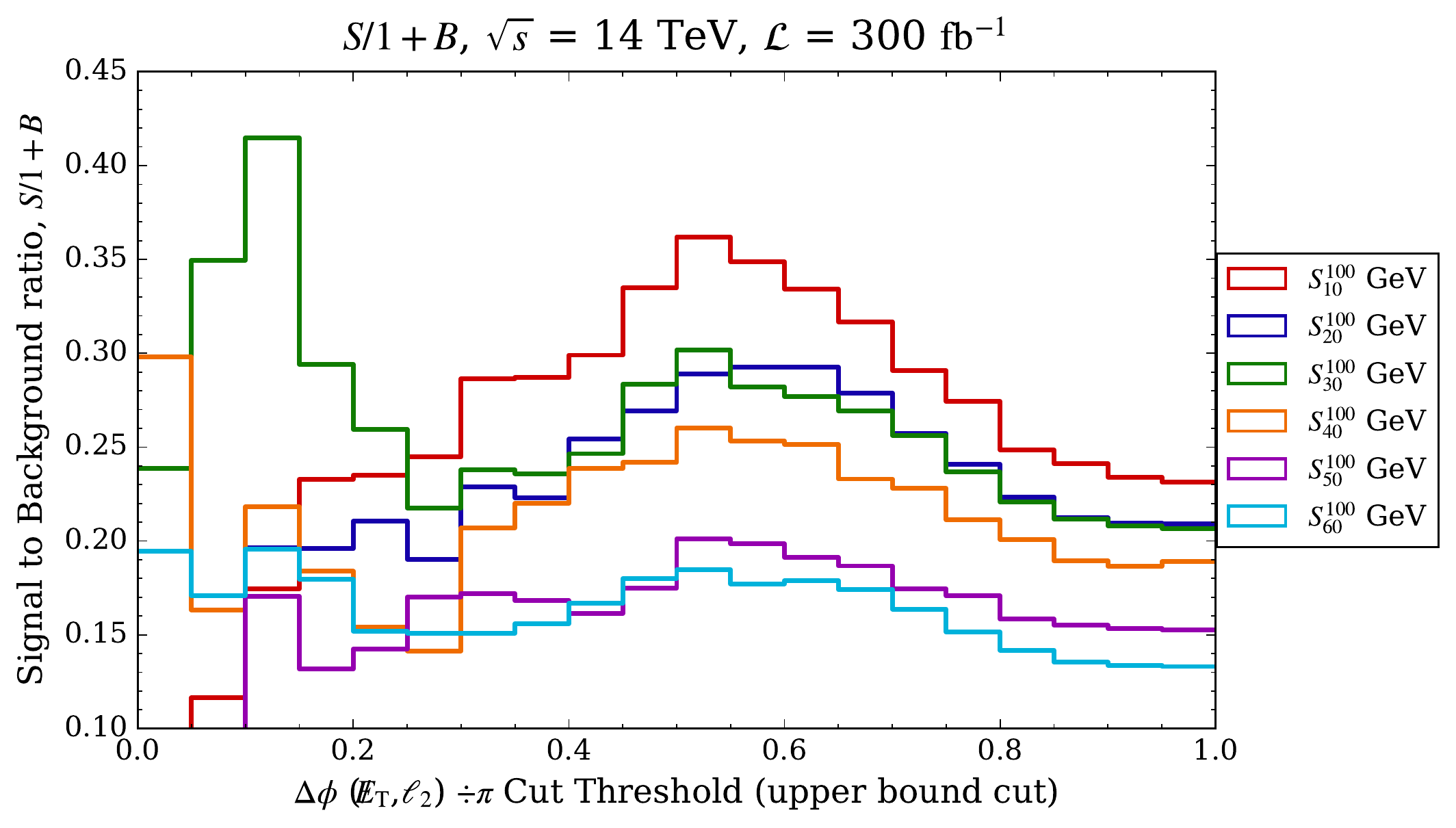}{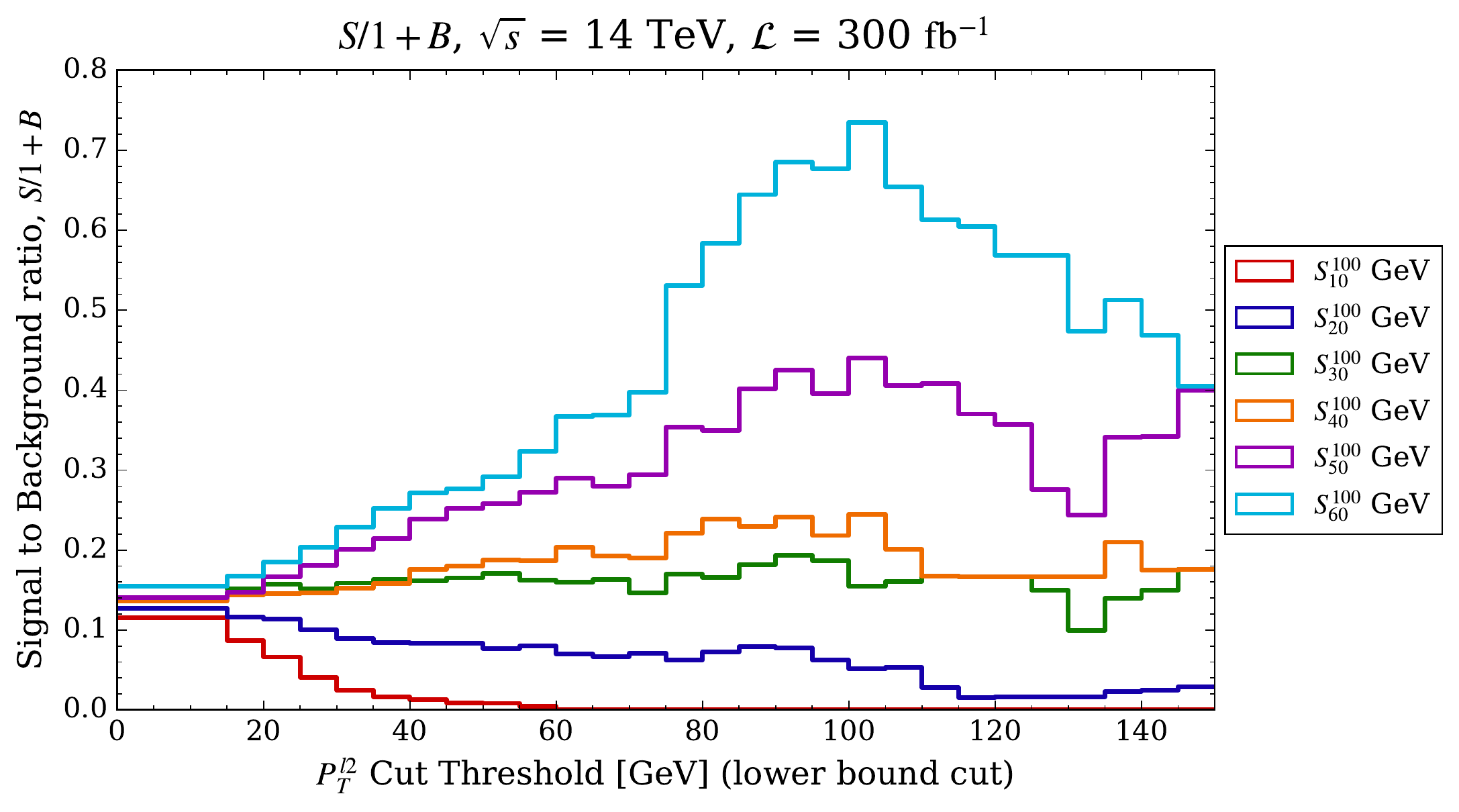}{fig:tertiary}
{
Signal-to-background ratios as a function of the final Table~\ref{tab:tertiary} cuts for
the intermediate and large mass gap scenarios.
The azimuthal angular separation variables, e.g. via an upper bound on $\Delta \phi (\met, \ell_2)$ (left),
prove useful for discriminating signal from background in the difficult intermediate mass splitting scenarios,
where leptonic transverse momentum magnitude becomes degenerate with that of vector boson decay products.
A lower bound on the minimal leptonic $P_T^{\ell_2}$ is useful for discriminating signal from background
when the mass gap is larger, and the slepton delivers a correspondingly enhanced boost to its decay products.
}

\subsection{$\Delta m=50,\,60\gev$}
For relatively large $\Delta m$, e.g., 50-60$\gev$, many of the prior considerations are reversed.
As the mass gaps become larger, the lepton plus invisible system becomes harder and less collimated than the background.
A better strategy to distinguish these scenarios from the others is to use lower bounds on the
angle $\Delta \phi (\ell_1, \met)/ \pi >0.25$ between the leading lepton and the missing transverse momentum vector,
and on the leptonic transverse momentum $p_T^{\ell_2} >40$~GeV (see the right panel of FIG.~\ref{fig:tertiary}).
Note that pushing the $P_T^{\ell_2}$ cut harder, say, to 80~GeV is beneficial for $\Delta m = 60$~GeV, but
is costly to the statistical significance for $\Delta m = 50$~GeV.
The signal significance after the application of all tertiary cuts can be above 4 for luminosity $300~{\rm fb}^{-1}$ and $S/B$ is above 25\%.

\subsection{Other Benchmarks}

As the smuon mass increases, the smuon production cross section drops. But, this is partially offset by increases
in some of the cut efficiencies due to the fact that there is more more $\met$ in the system.
In Table~\ref{tab:scaling} we show the significance for
luminosities of 300, 1000 and 3000 fb$^{-1}$ for smuon masses up to 300 GeV, and for mass gaps of 10-60 GeV.
Most of the benchmarks up to 200~GeV are able to be excluded at or near the 2-$\sigma$ level,
with signal-to-background on the order of 10\%.  For benchmarks as heavy as 300~GeV, the cross section
suppression is too large to overcome.
In Table~\ref{tab:5GeV} we examine improvements to the signal yields for small mass splittings
when the muon sensitivity is reduced from 10~GeV to 5~GeV. We notice that by reducing the threshold of muon $p_T$ enhances signal yields by $30 \%$ for $\Delta m \lesssim 20$ GeV scenarios.

\bgroup
\def\arraystretch{1.3}
\begin{table}[!htp]
\caption{Heavier model benchmarks $S^{m_{\tilde \mu}}_{\Delta m}$ for the small ($\Delta m = 10,20$~GeV),
intermediate ($\Delta m = 30,40$~GeV),
and large ($\Delta m = 50,60$~GeV),
mass gap tertiary event selection cuts.}
\label{tab:scaling}
\begin{center}
\begin{tabular}{|c||c|c||c|c||c|c|}
\hline
Benchmark & $S^{160}_{10}$ & $S^{160}_{20}$ & $S^{160}_{30}$ & $S^{160}_{40}$ & $S^{160}_{50}$ & $S^{160}_{60}$ \\
\hline
Events at $\mathcal{L}=300~{\rm fb}^{-1}$& 42 &  39 &  25 &  28 &  29 &  28 \\
\hline
$S \div (1+B)$& 0.24 &  0.22 &  0.11 &  0.12 &  0.12 &  0.12 \\
\hline
$S \div \sqrt{1+B}$& 3.2 &  2.9 &  1.7 &  1.9 &  1.9 &  1.8 \\
\hline
\hline
Benchmark & $S^{200}_{10}$ & $S^{200}_{20}$ & $S^{200}_{30}$ & $S^{200}_{40}$ & $S^{200}_{50}$ & $S^{200}_{60}$ \\
\hline
Events at $\mathcal{L}=1000~{\rm fb}^{-1}$& 72 &  67 &  42 & 46 &  53 &  64 \\
\hline
$S \div (1+B)$ & 0.12 &  0.11 & 0.06 & 0.06 & 0.07 & 0.08 \\
\hline
$S \div \sqrt{1+B}$ &  3.0 &  2.8 & 1.5 & 1.7 & 1.9 & 2.3 \\
\hline
\hline
Benchmark & $S^{300}_{10}$ & $S^{300}_{20}$ & $S^{300}_{30}$ & $S^{300}_{40}$ & $S^{300}_{50}$ & $S^{300}_{60}$ \\
\hline
Events at $\mathcal{L}=3000~{\rm fb}^{-1}$& 48 &  54 & 33 & 33 & 48 & 60 \\
\hline
$S \div (1+B)$& 0.03 & 0.03 & 0.01 & 0.01 & 0.02 & 0.03 \\
\hline
$S \div \sqrt{1+B}$& 1.1 &  1.3 & 0.7 & 0.7 & 1.0 & 1.2 \\
\hline
\end{tabular}
\end{center}
\end{table}
\egroup

\bgroup
\def\arraystretch{1.3}
\begin{table}[!htp]
\caption{Scaling of the $m_{\tilde \mu} = 110~\gev$ benchmark model with reduction of the muon reconstruction $p_T$ threshold to 5~GeV,
assuming small mass gap tertiary event selection cuts.}
\label{tab:5GeV}
\begin{center}
\begin{tabular}{|c||c|c|c|}
\hline
5~GeV muon & $S^{110}_{10}$ & $S^{110}_{20}$ & $S^{110}_{30}$ \\
\hline
Events at $\mathcal{L}=300~{\rm fb}^{-1}$ & 153 & 91 & 51 \\
\hline
Ratio to $P_T^\mu > 10$~GeV & 1.3 & 1.3& 1.0 \\
\hline
\end{tabular}
\end{center}
\end{table}
\egroup

\section{Conclusions}

We have considered strategies for detecting new physics scenarios in which the
new light particles are a bino-like LSP and a light slepton with a squeezed
spectrum ($\Delta m \lesssim 60\gev$).
It is well-known that
this scenario is difficult to probe, because the squeezed spectrum of new particles
implies that both the leptons and missing transverse energy produced through slepton pair
production are soft.  It is also well-known that one can improve detection prospects for this
scenario by requiring the emission of an additional jet, which gives the remaining system a transverse
boost; one then searches for a single hard jet, a same-flavor, opposite-sign lepton pair, and missing transverse energy.
We study enhancements to this strategy which can improve both signal significance and
the signal-to-background ratio, particularly focused on cuts which can distinguish the energy and
angular distribution of the signal events from SM background events. One can utilize the angular separation of the leptons from
each other and from the missing transverse energy and the  angular separation between the jet and the missing transverse energy
to distinguish signal from background events.

Focusing on the specific case $m_{\tilde \mu} = 110~\gev$, we found that these cuts can be used to achieve a larger signal-to-background ratio in the
small mass-splitting regime ($\Delta m = 10, 20~\gev$) than was found in previous analyses using alternative
cut strategies, while still maintaining discovery-level signal significance with $300~\fb^{-1}$ integrated luminosity.
We also found selection cuts which allow $4-5\sigma$ evidence to be found, with the same luminosity, for
$\Delta m$ as large as $60~\gev$.
The larger $\Delta m$ regions can be distinguished by using lower bounds on the $p_T$ of the leptons.
These selection cuts remain effective for larger choices of $m_{\tilde \mu}$,
up to and above $m_{\tilde \mu} \simeq 200~\gev$. We found that the LHC can set $\sim 1.5-3 \sigma$ exclusion limits on $\m_{\tilde\mu} \approx 200$ GeV, for $\Delta m \lesssim 60$ GeV with 1000 fb$^{-1}$ of integrated luminosity.

Although we have framed our study in terms of a MSSM scenario with a light bino-slepton pair, this
analysis can be readily adapted to other scenarios in which the LHC can pair-produce a lepton-partner,
which then decays to a nearly-degenerate invisible particle and a soft lepton.  However, a key feature of our
analysis is the use of angular distributions to distinguish slepton production from $W^+ W^-$ production,
a background which produces distinctive correlations in the outgoing leptons.  In a scenario in which the
lepton partner were also spin-1 instead of spin-0, one might reevaluate certain aspects of this strategy
(in particular, the selection on $\cos \theta^*_{\ell_1 \ell_2}$).

Finally, we note that we have only considered the case in which a bino-like LSP is nearly degenerate with
a slepton.  There has been great interest in case where the LSP is nearly degenerate with a squark, with
much work on LHC strategies for probing this scenario.  It would be interesting to determine if any of the
strategies we have discussed here could be adapted for that purpose.  In a similar vein, it would be interesting
to study if ILC searches could provide complimentary probes of scenarios with squeezed bino-slepton  spectra.

{\bf Acknowledgements}

We are grateful to Jamie Gainer and Xerxes Tata  for useful discussions.
B.~Dutta acknowledges support from DOE Grant DE-FG02-13ER42020.
K.~Fantahun, A.~Fernando, and J.~W.~Walker acknowledge support from NSF grant PHY-1521105.
T.~Ghosh is in part supported by the United States Department of Energy Grant Number de-sc 0016013.
J.~Kumar is supported in part by NSF CAREER grant PHY-1250573.
P.~Sandick is supported in part by NSF grant PHY-1417367.
P.~Stengel is supported in part by DOE grant DE-SC007859.
Part of this work was performed the Aspen Center for Physics, which is
supported by National Science Foundation grant PHY-1607611.
We are grateful to CETUP* and Oggie's Sports Bar, where part of this work was performed, for their hospitality.


\end{document}